\documentclass[11pt]{article}
\usepackage{amsmath, amsfonts, amssymb, amsthm}
\usepackage{graphicx, color}
\usepackage{latexsym}
\usepackage{epsfig}
\usepackage{setspace}

\newcommand{\mE}{\mathrm{E}}

\title{Modeling and Computation of High Efficiency and Efficacy Multi-Step Batch 
  Testing for Infectious Diseases}
\author{Hongshik Ahn, Haoran Jiang$^{*}$ and Xiaolin Li
        \vspace{.1in} \\
        Department of Applied Mathematics and Statistics  \\
        Stony Brook University \\
	Stony Brook, NY 11794-3600 \\
        $^{*}$Corresponding author. email: haoran.jiang@stonybrook.edu}
\date{}

\begin{document}
\maketitle
\setcounter{page}{1}
\begin{abstract}
We propose a mathematical model based on probability theory to optimize COVID-19
testing by a multi-step batch testing approach with variable batch sizes. This
model and simulation tool dramatically increase the efficiency and efficacy of
the tests in a large population at a low cost, particularly when the infection
rate is low. The proposed method combines statistical modeling with numerical
methods to solve nonlinear equations and obtain optimal batch sizes at each step
of tests, with the flexibility to incorporate geographic and demographic
information. In theory, this method substantially improves the false positive
rate and positive predictive value as well. We also conducted a Monte Carlo
simulation to verify this theory. Our simulation results show that our method
significantly reduces the false negative rate. More accurate assessment can be
made if the dilution effect or other practical factors are taken into consideration.
The proposed method will be particularly useful for the early detection of infectious
diseases and prevention of future pandemics.  The proposed work will have broader
impacts on medical testing for contagious diseases in general.

\end{abstract}

\section{Introduction}
\label{sec:intro}

To fight the COVID-19 pandemic with limited resources, batch tests were
recommended by pooling multiple swab samples.
Samples from different individuals are pooled into one batch and then a
high-throughput PCR test is conducted (Cheng, 2020;
Lohse et al., 2020; Shani-Narkiss et al., 2020). If the batch is tested
negative, then it can be deduced that all samples were negative.
Otherwise, each sample needs to be tested individually. Pooling data was
originally proposed by Dorfman (1943) for detecting syphilis in US soldiers
during World War II. Farrington (1992), Gastwirth and Hammick (1989),
Chen and Swallow (1990), Gastwirth and Johnson (1994), Hardwick et al.\
(1998), Hung and Swallow (2000), Vansteelandt et al.\ (2000), Xie (2001),
Bilder and Tebbs (2009) and Chen et al.\ (2009) developed group-testing
regression on parametric models. Delaigle and Meister (2011) developed a
nonparametric method to estimate the conditional probability of contamination
for pooled data. Wang, McMahan et al.\ (2014) developed semiparametric group
testing regression models. France et al.\ (2015) used pooling samples to
reduce the number of tests required for detection of anthrax spores. Kline
et al.\ (1989), Behets et al.\ (1990), Lindan et al.\
(2005), Pilcher et al.\ (2005), Hourfar et al.\ (2008), Stramer et al.\ (2013)
and Warasi et al.\ (2016) developed group testing methods to screen sexually
transmitted diseases. Pooled samples were used to detect other infectious
diseases in Busch et al.\ (2005), Van et al.\ (2012) and Wang, Han et al.\ (2014).
Nagi and Raggi (1972), Fahey et al.\ (2006), Wahed et al.\ (2006) and Lennon
(2007) pooled observations to test for contamination by a toxic substance.
Huang (2009) and Huang and Tebbs (2009) studied measurement error models for
group testing data.

By grouping individuals, batch testing significantly reduces the number of
tests, providing an efficient method to detect community transmission
(Hogan et al., 2020). Batch testing has become more relevant recently, as
state and local governments seek to test as many people as possible to
transition safely back to normal life. In July 2020, the US Food and Drug
Administration issued emergency authorization for sample pooling in diagnostic
testing (US Food and Drug Administration, 2020). Places abroad such as South
Korea have been using pooling methods to sample in batches of a fixed size in
high-risk communities (Park and Koo, 2020; Kwak, 2020; Korea Center for Disease
Control \& Prevention, 2020). However, the drawback to batch testing is a much
higher false negative rate than individual testing.

In this study, we introduce a batch-based approach which simultaneously
addresses the problem of limited resources and testing accuracy. We
consider a multi-step testing procedure with variable batch sizes where
each step divides the population into subpopulations based on the
previous step's results. We introduce a method to estimate the optimal batch
sizes given the infection rates of subpopulations to efficiently and
accurately test entire population. The proposed
method incorporates testing errors and optimizes batch sizes at each step
and for each subpopulation. Shani-Narkiss et al.\ (2020) also considered
a multi-step testing procedure with variable batch sizes. However, their
approach assumes no testing errors, and their batch sizes are limited to
powers of two.

For batch testing, typically a sample from each person is divided into
multiple aliquots for separate tests (Mutesa et al., 2020).
The population is split into subpopulations with negative test results
(batch negative) and positive test results (batch positive). Each
subpopulation is given another round of batch tests where the batch size
increases for the batch negatives and decreases for the batch positives.
We iterate this procedure on each subpopulation, where at each step we
can estimate the infection rate. This process is continued until one of
the following conditions is satisfied: (i) the process results in three
batch negatives or three batch positives, (ii) the infection rate of the
subpopulation becomes higher than 30\%, (iii) the optimal batch size is
reduced to 2. The samples are randomly assigned to different batches with
the size determined by the infection rate in each round.
Batch testing requires more tests than individual testing if the infection
rate is over 30\% with batch of size less than 3 (Armend{\'a}riz et al., 2020).
Details of this procedure are given in Section~\ref{sec:proc}.
To apply our approach most effectively, we can first divide the population
based on infection rates, for example by dividing based on geography,
population density, proximity to highly infected regions, etc. Information
given by various methods including mobile apps and online mapping
(Lee and Lee, 2020) may help track the virus and divide the
population into different groups.

We also address the efficacy of the tests. The false negative rate of the
COVID-19 PCR test for an individual is known to be near 15\%, ranging from
10 to 30\% (Xiao et al., 2020; West et al., 2020; Yang and Yan, 2020). By
one-step batch testing, the false negative rate increases (Shuren, 2020).
By our multi-step batch testing procedure, the false negative rate is
substantially reduced. To the best of our knowledge, no other studies have
attempted estimating optimal batch sizes by taking testing errors into
account. We also have derived the sensitivity and specificity of the
proposed multi-step batch testing. The dilution effect in batch testing
(Hwang, 1976; McMahan et al., 2013; Yelin et al., 2020) is not considered
in this study. The false negative rate may be increased by the dilution effect.

If three batch negatives occur before getting three batch positives, then we
conclude that the individuals in the batch of the final round are not infected.
For people whose samples result in three batch positives before three batch
negatives, each sample needs to be tested individually to find out which was
positive. To reduce the false negative rate, up to three individual tests are
performed for the samples from each individual in this group. For a
population of size 100,000 with an infection rate of .1\%, if the false
negative rate is 15\% and false positive rate is 1\% for an individual test,
then our method requires approximately 7,000 tests to test the entire population.
Results for different infection rates are detailed in Section~\ref{sec:simul}.
Our method reduces the false negative rate
to approximately 3\%, and the false positive rate to near zero.

It is a well-known fact that the positive predictive value (PPV) is very low
when the infection rate is low even if the sensitivity is very high. For a
false positive rate of 1\% and false negative rate of 15\%, the PPV of an
individual test is 8\% for an infection rate of .1\%, and 46\% for an
infection rate of 1\%. According to our simulation, our multi-step batch
testing procedure improves the PPV to 89\% and 93\%, respectively.
This is because individual tests are conducted on the subject in the final
positive batches which have higher infection rates than the entire population
(see Section~\ref{sec:ppvnpv} and Section~\ref{sec:simul}).

The proposed batch testing procedure is compared with matrix pool testing as
well as single batch testing with fixed or variable batch sizes. Two dimensional
pooling strategy has been studied by researchers including Barillot et al.\ (1991),
Amemiya et al.\ (1992), Phatarfod and Sudbury (1994) and Hudgens and Kim (2011).
The false negative rate decreases to 1\% if $12 \times 12$ matrix pool tests are
conducted 3 times parallel with up to 3 sequential individual tests to all
the positive crossings. However, the number of tests to cover the whole population
is much more than that of the proposed multi-step batch testing procedure, and it
is more than a half the population size even when the infection rate of the
population is .1\% (see Section~\ref{sec:simul}).

The original purpose for batch testing was to prevent the spread of disease in
high-risk communities by testing everyone, symptomatic or not.  However, the
proposed method can be applied to the general population, due to its flexibility
in dealing with various infection rates and its substantially greater
sensitivity compared to individual testing and current batch testing approaches.
More specifically, our method can accurately test a large population with
limited resources by dividing the population based on infection rates, for
example based on geography, population density, proximity to highly infected
regions, etc. The proposed method is particularly efficient when the infection
rate is low. This method can be applied to effectively combat a future pandemic
of new diseases.

\section{Mathematical Model}
\subsection{Optimal batch size assuming no testing errors}
\label{sec:binprob}

We begin by studying a simple case where the accuracy of a virus screening test
is 100\%. Suppose that the infection rate in a population is $p$ (rate of no
infection is $q=1-p$). Let $X$ be a random variable denoting the number of positive cases in a
batch of size $n$. Then $X$ follows a binomial distribution with $n$ trials
and success rate $p$. The probability of $k$ positive cases in the batch is
\begin{equation}
 P(X=k) = \binom{n}{k} p^k q^{n-k}, ~ k=0, \cdots, n
 \label{eq:binprob}
\end{equation}
and the probability that the batch is tested negative is $P(X=0)=q^n$.

Using an initial guess of the batch size $n$, we want to estimate $q$.
Let $A=P(X=0)$. Then after a good number of batch tests, we can estimate $q$ by
\[ A \equiv q^n \Longrightarrow q=A^{1/n}. \]
Suppose we test a population in batches of size $n$, then test all individuals
who were in positively tested batches. For a population of size $N$, the expected
total number of tests to be performed to identify all positive carriers is
(see also Armend{\'a}riz et al., 2020)
\begin{equation}
 T(n) = \frac{N}{n} + n \left( 1-q^n \right) \frac{N}{n}
   = \frac{N}{n} + N \left( 1-q^n \right) = N \left( \frac{1}{n} +1-q^n \right).
   \label{eq:opt1}
\end{equation}
To find the optimal $n$, we minimize $T(n)$ as
\[ T'(n) = N \left( -\frac{1}{n^2} - q^n \ln q \right) = 0
  \Longrightarrow n^2 q^n \ln q = -1 \]
We can solve this equation for $n$ numerically.

\subsection{Optimal batch size given testing errors}
\label{sec:model}

We conduct a hypothesis testing with null hypothesis $H_0$: The subject is uninfected versus
alternative hypothesis $H_1$: The subject is infected. Denote the probability of a type I
error (false positive rate) as $\alpha$ and the probability of a type II error (false
negative rate) as $\beta$. For multiple rounds of batch testing, let the infection rate be
$p_1$ and $q_1=1-p_1$ in the initial batch tests for
the whole population, and $X$ the random variable denoting the number of positive cases in
a batch of size $n_1$. Then the probability of a batch negative based on (\ref{eq:binprob})
can be obtained as
\begin{equation}
 (1-\alpha) P(X=0) + \beta P(X > 0) = (1-\alpha) q_1^{n_1} + \beta \left( 1-q_1^{n_1} \right)
  = (1-\alpha-\beta) q_1^{n_1} + \beta.
 \label{eq:probn1}
\end{equation}
Using an initial guess of the batch size $n_1$, we want to estimate $q_1$.
After a good number of batch tests, we can estimate $q_1$ by
\begin{equation}
 A_1 \equiv (1-\alpha-\beta) q_1^{n_1} + \beta \Longrightarrow
  q_1 = \left( \frac{A_1-\beta}{1-\alpha-\beta} \right)^{1/n_1}.
 \label{eq:A1}
\end{equation}

Suppose all the subjects in the positive batches get individual tests.
If the size of the entire population is $N_1$, the required number of tests can be
obtained from (\ref{eq:opt1}) by substituting (\ref{eq:probn1}) for $q^n$ as
\begin{equation}
 T_1(n_1) = N_1 \left[ \frac{1}{n_1} + 1-A_1 \right]
  = N_1 \left[ \frac{1}{n_1}+1-\beta - (1-\alpha-\beta)q_1^{n_1} \right].
   \label{eq:optn1}
\end{equation}
To find the optimal batch size $x$, we minimize $T_1 (x)$ as
\[ T_1'(x) = N_1 \left[ -\frac{1}{x^2} - (1-\alpha-\beta) q_1^x \ln q_1 \right] = 0
  \Longrightarrow x q_1^{x/2} = [-(1-\alpha-\beta) \ln q_1]^{-1/2}. \]
This can be solved for $x$ numerically.
In this study, we used the secant method to solve the equation. The optimal batch size
$n_1$ is either $\mbox{floor}(x)$ or $\mbox{ceiling}(x)$ which has the lower value of
$T(\cdot)$. The initial batch tests can be conducted using $n_1$ for the whole population.

Continuing this process, after the $(i-1)$th round, the subjects have been divided into
subpopulations. Each subpopulation was batch tested with batch sizes determined from
that round. Fixing one of these subpopulations, we now split it into two smaller
subpopulations for the $i$th round, according to the batch test results from the
$(i-1)$th round. Let $N_i$ denote the number of subjects in population belonging
to test-negative batches from the $(i-1)$th round, and $p_i$ denote the infection rate
of this subpopulation. The expected size of the subpopulation in the $i$th round of
batch tests is
\begin{equation}
 N_i = N_{i-1} \left[ (1-\alpha-\beta) q_{i-1}^{n_{i-1}} + \beta \right].
 \label{eq:Ni}
\end{equation}
In this subpopulation, the probability that at least one subject in a batch is infected is
\begin{equation}
 r_i \equiv \frac{\beta \left( 1-q_{i-1}^{n_{i-1}} \right)}{(1-\alpha) q_{i-1}^{n_{i-1}}
   + \beta \left( 1-q_{i-1}^{n_{i-1}} \right)}
 \label{eq:ri}
\end{equation}
and the probability that all the subjects in a batch are not infected is $1-r_i$.
The estimated number of infected subjects in this subpopulation is
\[ m_i \equiv \frac{N_i}{n_{i-1}} \cdot r_i \cdot \frac{\mE (X_{i-1})}{P(X_{i-1} > 0)}
  = \frac{N_i}{n_{i-1}} \cdot r_i \cdot \frac{n_{i-1} p_{i-1}}{1-q_{i-1}^{n_{i-1}}}
  = \frac{N_i p_{i-1} r_i}{1-q_{i-1}^{n_{i-1}}}. \]
Here, $\mE (X_{i-1})/P(X_{i-1}>0)$ is the expected number of infected subjects in a
test-positive batch in the $(i-1)$th round, where $X_{i-1}$ is a binomial random variable
with $n_{i-1}$ trials and success rate $p_{i-1}$. Therefore, the infection rate
\begin{equation}
 p_i= \frac{m_i}{N_i} = \frac{p_{i-1} r_i}{1-q_{i-1}^{n_{i-1}}}
   \label{eq:pi}
\end{equation}
of the subpopulation in the $i$th round can be used for estimating the optimal batch size
and the required number of tests. The optimal number of required tests can be obtained by
replacing $n_1$ with $n_i$ and $q_1$ with $q_i$ in (\ref{eq:probn1}) and (\ref{eq:optn1})
to get
\begin{equation}
 T_i(n_i) = N_i \left[ \frac{1}{n_i}+1-\beta - (1-\alpha-\beta)q_i^{n_i} \right]
 \label{eq:optni}
\end{equation}
and the optimal batch size for this round is obtained by minimizing $T_i(\cdot)$.

For example, suppose the estimated infection rate is $p_1=.01$ and $\alpha=.01, ~ \beta=.15$
for a population of size 100,000. Then $T_1(x)$ in (\ref{eq:optn1}) is minimized when $n_1=12$.
For the second round, the expected subpopulation size is obtained from (\ref{eq:Ni}) as
$N_2 = 100,000 \left[ .84 (.99)^{12}+.15 \right] =$ 89,456.
From (\ref{eq:ri}), $r_2 = .15 \left[1-(.99)^{12} \right]/\left\{(1-.01) (.99)^{12}
+ 0.15 \left[ 1-(.99)^{12} \right] \right\} = .019$ and the estimated infection rate in this
subpopulation is obtained from (\ref{eq:pi}) as $p_2 = (.01) (.019)/\left[1-(.99)^{12} \right]
= 0.00167$. For this infection rate, the optimal batch size $n_2=27$ is obtained by minimizing
$T_2 (x)$ in (\ref{eq:optni}). These numbers can be found in the second part of
Table~\ref{tab:proc} under Round~2.

Further, we can estimate the optimal batch size for the subpopulation consisting of batch positives.
In the $(i-1)$th round, the probability that a batch of size $n_{i-1}$ is tested positive is
\[ \alpha P(X=0) + (1-\beta) P(X>0) = \alpha q^{n_{i-1}} + (1-\beta) \left( 1-q^{n_{i-1}} \right)
  = (1-\beta) - (1-\alpha-\beta) q^{n_{i-1}}. \]
The size of this subpopulation can be obtained by modifying (\ref{eq:Ni}) as
\[ N_i = N_{i-1} \left[ (1-\beta)-(1-\alpha-\beta) q_{i-1}^{n_{i-1}} \right]. \]
In this subpopulation, the probability that at least one subject in a batch is infected is
\[ r_i \equiv \frac{(1-\beta) \left( 1-q^{n_{i-1}} \right)}{\alpha q^{n_{i-1}}
  + (1-\beta) \left( 1-q^{n_{i-1}} \right)} \]
and the probability that all the subjects are not infected is $1-r_i$.
The estimated infection rate $p_i$ in this subpopulation can be obtained as (\ref{eq:pi}).
This infection rate can be used for estimating the optimal batch size $n_i$ and the required
number of tests for the subpopulation in the $i$th round. The accuracy of these formulae
developed in this section has been confirmed by the simulation
results given in Section~\ref{sec:simul}.

\begin{figure}
\centerline{\includegraphics[height=3.3in]{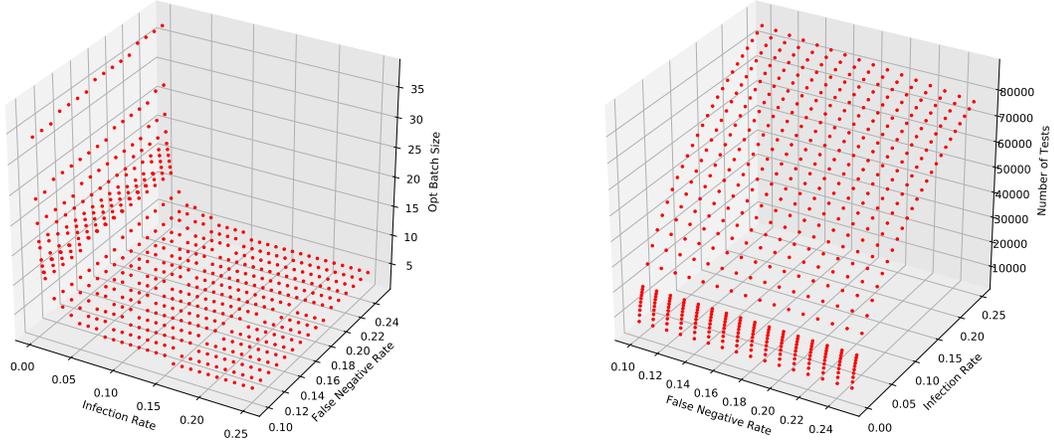}}
\vspace{-.3in}
\caption{Optimal batch size (left) and required number of batches (right) for the first round of
 batch tests. False positive rate: $.01$, population size: 100,000.
  \label{fig:optn}}
\end{figure}

Figure~\ref{fig:optn} displays the optimal batch size and the required number of batches for a
population of size 100,000 as the infection rate ranges from .001 to .25 for the first round,
and the false negative rate ranges from .1 to .25. The false positive rate is fixed at $.01$.
Table~\ref{tab:opt} displays the optimal batch sizes for different infection rates (a) when no
testing errors are assumed and (b) when the false negative rate is 15\% and the false positive
rate is 1\%.

\begin{table}[tbp]
 \centering
 \caption{Optimal batch size (a) when there are no testing errors, and (b) when the false
  negative rate is .15, false positive rate is .01 for individual tests.}
 \label{tab:opt}
\vspace{1em}
\begin{tabular}{ccccccccccccc}
\hline
Infection rate & .001 & .002 & .003 & .004 & .005 & .006 & .007 & .008 & .009 &.01 & .02 & .03 \\
 \cline{2-13}
 (a) without error & 32 & 23 & 19 & 16 & 15 & 13 & 12 & 12 & 11 & 11 & 8 & 6 \\
 (b) with error & 35 & 25 & 21 & 18 & 16 & 15 & 14 & 13 & 12 & 12 & 8 & 7 \\
\hline
 Infection rate & .04 & .05 & .06 & .07 & .08 & \multicolumn{2}{c}{.09 to .12} & \multicolumn{2}{c}{.13 to .17}
     & \multicolumn{2}{c}{.18 to .25} & \\
 \cline{2-13}
 (a) without error & 6 & 5 & 5 & 4 & 4 & \multicolumn{2}{c}{4} & \multicolumn{2}{c}{3} & \multicolumn{2}{c}{3} & \\
 (b) with error & 6 & 6 & 5 & 5 & 5 & \multicolumn{2}{c}{4} & \multicolumn{2}{c}{4} & \multicolumn{2}{c}{3} & \\
\hline
\end{tabular}
\end{table}

\subsection{Sensitivity and specificity of a batch test}

Let us consider one-step batch testing. If a batch is tested negative, then we conclude
that all samples in the batch are negative. If a batch is tested positive, then each sample
in the batch needs to be tested individually. Let the binomial random variable $X$ denote
the number of positive cases in a batch of size $n$ with an infection rate of $p$. We
continue to denote $\alpha$ as the probability of a type I error and $\beta$ as the
probability of a type II error for an individual test.
Table~\ref{tab:mtx} displays the probabilities for the confusion matrix.

\begin{table}[tbp]
 \centering
 \caption{Probability in each cell of the confusion matrix for single batch testing.}
 \label{tab:mtx}
\vspace{.5em}
 \begin{tabular}{ccc|c}
        &     & \multicolumn{2}{c}{True condition} \\
        &     & No samples are infected & At least one sample is infected \\
        \cline{3-4}
 Test   & $-$ & (a) $(1-\alpha) P(X=0)$ & (b) $\beta P(X>0)$ \\
        \cline{3-4}
 Result & $+$ & (c) $\alpha P(X=0)$ & (d) $(1-\beta) P(X>0)$ \\
        \cline{3-4} \\
 \end{tabular}
\end{table}

\subsubsection{Sensitivity}
\label{sec:sens}

The probability that at least one sample is infected in each batch is $P(X>0)$ which is
the sum of the probabilities in cells (b) and (d). In cell (b), it can be deduced that all
the samples are negative, and no more tests are given. In cell (d), each sample gets an
individual test, and the probability of false negative here is $\beta (1-\beta)P(X>0)$.
Thus, the false negative rate is
\[ \frac{\beta P(X>0) + \beta (1-\beta)P(X>0)}{P(X>0)} = \beta + \beta (1-\beta)
     =\beta (2-\beta). \]

For example, if $\beta=.15$, then the false negative rate of batch testing is $.15 (2-.15)=.2775$,
and thus the sensitivity is .7225. For $\beta =$ .1, .2 and .25, the sensitivity of batch
testing is .81, .64 and .5625, respectively. Note that the sensitivity of
a batch test depends on neither the infection rate nor the batch size. The above result is
supported by our simulation given in Section~\ref{sec:simul}. It confirms that
the sensitivity is decreased by conventional batch testing. In contrast, our multi-step batch
testing method has a significantly higher sensitivity than conventional individual tests as
well as single batch testing as shown in Section~\ref{sec:proc}.

\subsubsection{Specificity}
\label{sec:spec}

The false positive error occurs in cells (c) and possibly in (d) in Table~\ref{tab:mtx},
and all the samples in these cells go through individual tests.
In cell (c), the probability that a sample is incorrectly tested positive
in the individual tests is
\begin{equation}
 \alpha^2 P(X=0)= \alpha^2(1-p)^n.
 \label{eq:spec1}
\end{equation}
In cell (d), the expected infection rate in each batch is $p/P(X>0)$.
The probability that an uninfected sample is incorrectly tested positive
in the individual tests in this cell is
\begin{equation}
 (1-\beta) P(X>0) \alpha \left[ 1- \frac{p}{P(X>0)} \right]
 = \alpha (1-\beta) [ P(X>0)-p]
 = \alpha (1-\beta) \left[ 1-(1-p)^n - p \right].
  \label{eq:spec2}
\end{equation}
The sum of (\ref{eq:spec1}) and (\ref{eq:spec2}) is
\[  \alpha^2 (1-p)^n + \alpha (1- \beta) \left[ 1-p - (1-p)^n \right]
  = \alpha (1-\beta) (1-p) + \alpha (\alpha+\beta-1) (1-p)^n. \]
Therefore, the false positive rate of batch testing is
\[ \mbox{Fp} \equiv \frac{\alpha (1-\beta) (1-p) +  \alpha (\alpha+\beta-1) (1-p)^n}{1-p}
 = \alpha (1-\beta) +  \alpha (\alpha+\beta-1) (1-p)^{n-1} \]
and the specificity of batch testing is
\begin{equation}
 1-\mbox{Fp} =  (1-\alpha + \alpha \beta) + \alpha ( 1 - \alpha - \beta ) (1-p)^{n-1}.
 \label{eq:spec}
\end{equation}

Unlike the sensitivity of batch testing, the batch size, infection rate, probability
of a type I error, and probability of a type II error contribute to (\ref{eq:spec}).
For $\alpha=$.01 \& .03, and $\beta=$.1, 15, .2 \& .25, the specificity of batch testing
(\ref{eq:spec}) using a fixed batch size of 10, and using the optimal batch
size is given in Table~\ref{tab:spec}. The specificity is substantially improved by batch
testing. These results closely match with our simulation results given in
Section~\ref{sec:simul}.

\begin{table}[tbp]
 \centering
 \caption{Specificity for single batch testing with fixed batch size of 10 and optimal
  batch size for the false positive rate of .01, .03, and the false negative rate of .1, .15, .2,
  .25 for individual tests.}
 \label{tab:spec}
\vspace{1em}
\begin{tabular}{cccc|ccccc}
\hline
$\alpha$ & $\beta$  &          & Infection rate      & .001  & .01   & .03   & .05   & .10 \\
 \hline
     &    & Batch size 10 & Specificity & .9998 & .9991 & .9978 & .9966 & .9944 \\
     \cline{3-9}
     & .1 & Optimal       & Batch size  & 34    & 11    & 7     & 5     & 4 \\
     &    & batch size    & Specificity & .9996 & .9990 & .9984 & .9982 & .9975 \\
    \cline{2-9}
     &    & Batch size 10 & Specificity & .9998 & .9992 & .9979 & .9968 & .9948 \\
     \cline{3-9}
     & .15 & Optimal      & Batch size  & 35    & 12    & 7     & 6     & 4 \\
 .01 &     & batch size   & Specificity & .9996 & .9990 & .9985 & .9980 & .9976 \\
    \cline{2-9}
     &    & Batch size 10 & Specificity & .9998 & .9992 & .9980 & .9970 & .9951 \\
     \cline{3-9}
     & .2 & Optimal       & Batch size  & 36    & 12    & 7     & 6     & 4 \\
     &    & batch size    & Specificity & .9996 & .9991 & .9986 & .9981 & .9978 \\
    \cline{2-9}
     &    & Batch size 10 & Specificity & .9998 & .9993 & .9981 & .9972 & .9954  \\
     \cline{3-9}
     & .25 & Optimal      & Batch size  & 37    & 12    & 7     & 6     & 5  \\
     &     & batch size   & Specificity & .9996 & .9991 & .9987 & .9982 & .9974  \\
\hline
     &    & Batch size 10 & Specificity & .9989 & .9968 & .9928 & .9895 & .9831 \\
     \cline{3-9}
     & .1 & Optimal       & Batch size  & 34    & 11    & 7     & 5     & 4 \\
     &    & batch size    & Specificity & .9983 & .9966 & .9947 & .9943 & .9920 \\
    \cline{2-9}
     &    & Batch size 10 & Specificity & .9989 & .9970 & .9932 & .9900 & .9840  \\
     \cline{3-9}
     & .15 & Optimal      & Batch size  & 36    & 12    & 7     & 6     & 4  \\
 .03 &     & batch size   & Specificity & .9983 & .9965 & .9950 & .9935 & .9924 \\
    \cline{2-9}
     &    & Batch size 10 & Specificity & .9989 & .9971 & .9936 & .9906 & .9849 \\
     \cline{3-9}
     & .2 & Optimal       & Batch size  & 37    & 12    & 7     & 6     & 4 \\
     &    & batch size    & Specificity & .9983 & .9967 & .9952 & .9939 & .9928 \\
    \cline{2-9}
     &    & Batch size 10 & Specificity & .9989 & .9972 & .9939 & .9911 & .9859 \\
     \cline{3-9}
     & .25 & Optimal      & Batch size  & 38    & 13    & 8     & 6     & 5  \\
     &     & batch size   & Specificity & .9983 & .9966 & .9950 & .9942 & .9917 \\
\hline
\end{tabular}
\end{table}

\subsection{PPV and NPV}
\label{sec:ppvnpv}

\begin{figure}
\centerline{\includegraphics[height=3in]{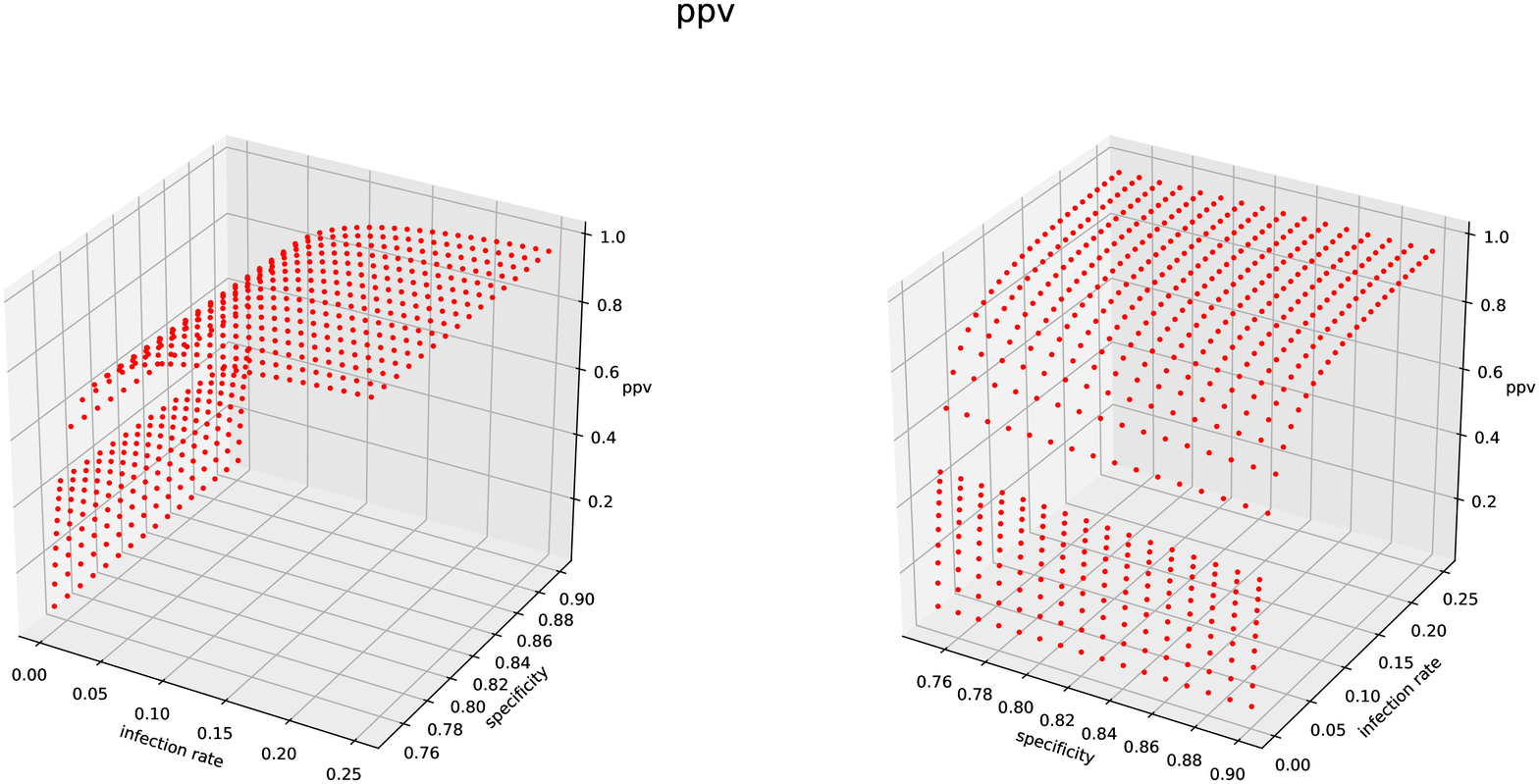}}
\centerline{\includegraphics[height=3in]{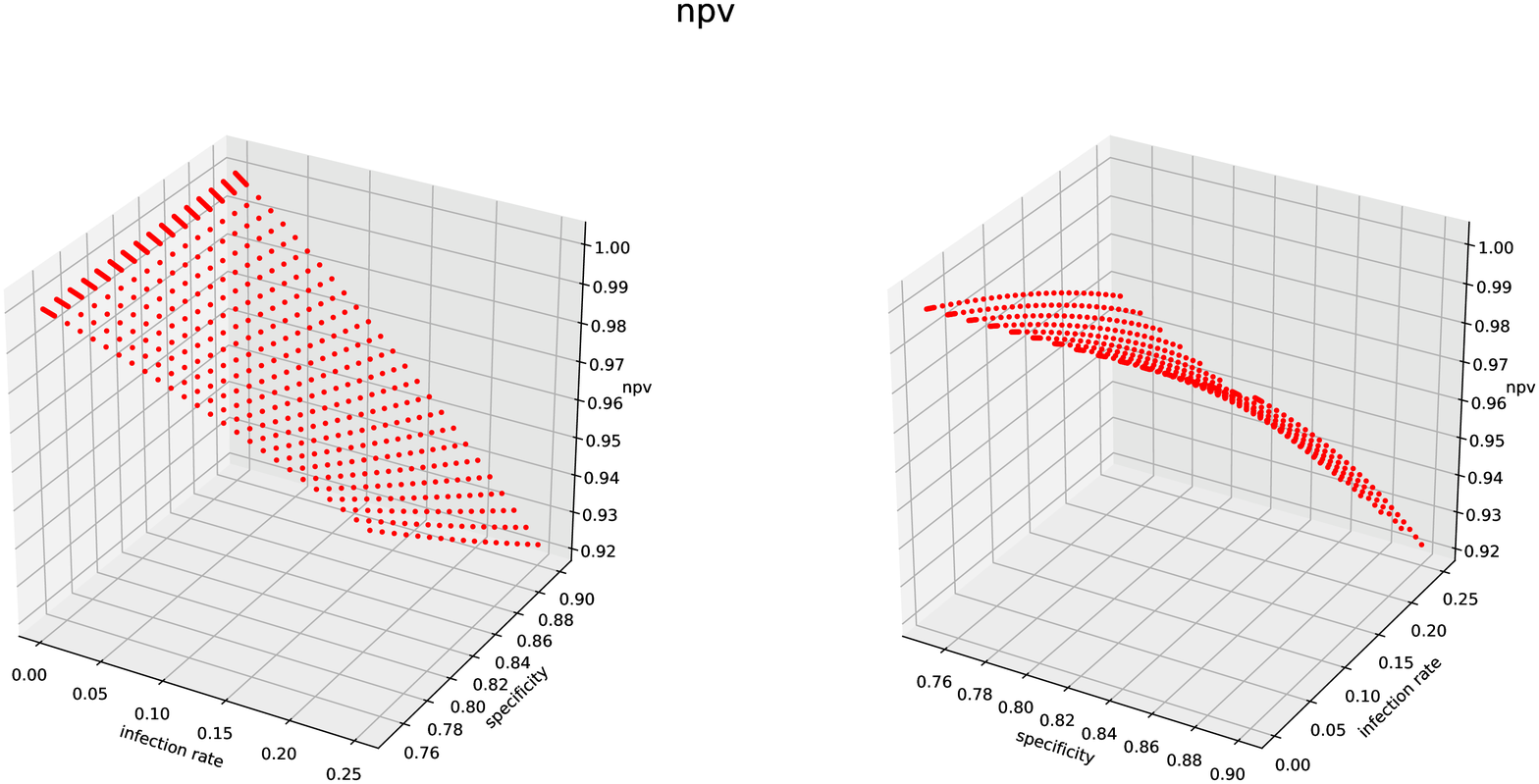}}
\caption{PPV (above) and NPV (below).
  \label{fig:ppvnpv}}
\end{figure}

Let us define $E$ as the event that an individual is infected by
the virus, and $B$ as the event that an individual got a positive test result. Then by the Bayes'
rule, PPV (positive predictive value: an individual is infected given a positive test result) is
\[ P(E|B) = \frac{P(B|E) P(E)}{P(B|E) P(E) + P \left( B | E^C \right) P \left( E^C \right)} \]
and NPV (negative predictive value: an individual is not infected given a negative test result) is
\[ P \left( E^C |B^C \right) = \frac{P \left( B^C |E^C \right) P \left( E^C \right)}
    {P \left( B^C |E^C \right) P \left( E^C \right) + P \left( B^C | E \right) P(E)}. \]
For example, if the infection rate in the population is 1\%, sensitivity
is 85\%, and specificity is 99\%, then $P(E)=0.01, ~ P(B|E)=0.85$ and
$P \left( B^C | E^C \right)=0.99$. Therefore, the PPV is
.4620 and the NPV is .9985.
If the infection rate is .1\%, then the PPV is .0078 and the NPV is .9998.
For the infection rates $p \in [.001, .2]$, sensitivity within the range
of [.75, .90] and the specificity of .99, Table~\ref{tab:trpos} illustrates PPV and NPV.
Figure~\ref{fig:ppvnpv} displays the PPV (above) and NPV (below).
By batch testing, the PPV is significantly improved.
If we denote $S_e(B)$ as the sensitivity and $S_p(B)$ as the specificity of single-step
batch testing, the PPV and NPV of single-step batch testing are given as
\begin{eqnarray*}
 \mbox{PPV}(B) &=& \frac{(1-p)[1-S_p(B)]}{(1-p)[1-S_p(T)]+p S_e(B)} \\
 \mbox{NPV}(B) &=& \frac{p[1-S_e(B)]}{p[1-S_e(T)]+(1-p) S_p(B)} \\
\end{eqnarray*}
respectively (Fletcher et al., 1988; Litvak et al., 1994; Kim et al., 2007). For the
infection rate of .1\%, PPV$(B)$ is .8049 and NPV$(B)$ is .9997. For the infection rate of
1\%, PPV$(B)$ is .8983 and NPV$(B)$ is .9972. These numbers are obtained by substituting the
values from Section~\ref{sec:sens} and Section~\ref{sec:spec} when a fixed batch size of 10
is used. The PPV is further improved by our multi-step batch testing procedure (see
Section~\ref{sec:simul}).

\begin{table}[tbp]
 \centering
 \caption{PPV and NPV.} \label{tab:trpos}
\vspace{1em}
\begin{tabular}{ccccccccccc}
\hline
 & & \multicolumn{9}{c}{Sensitivity$^{(1)}$} \\
    \cline{3-11}
$p^{(2)}$ & & .75 & .77 & .79 & .81 & .83 & .85 & .87 & .89 & .90 \\
\hline
.001 & PPV & .0070 & .0072 & .0073 & .0075 & .0077 & .0078 & .0080 & .0082 & .0083 \\
     & NPV & .9997 & .9998 & .9998 & .9998 & .9998 & .9998 & .9999 & .9999 & .9999 \\
    \cline{2-11}
.002 & PPV & .1307 & .1337 & .1367 & .1397 & .1426 & .1455 & .1485 & .1514 & .1528  \\
     & NPV & .9995 & .9995 & .9996 & .9996 & .9997 & .9997 & .9997 & .9998 & .9998 \\
    \cline{2-11}
.004 & PPV & .2315 & .2362 & .2409 & .2455 & .2500 & .2545 & .2589 & .2633 & .2655 \\
     & NPV & .9990 & .9991 & .9991 & .9992 & .9993 & .9994 & .9995 & .9996 & .9996 \\
    \cline{2-11}
.006 & PPV & .3116 & .3173 & .3229 & .3284 & .3338 & .3391 & .3443 & .3495 & .3520 \\
     & NPV & .9985 & .9986 & .9987 & .9988 & .9990 & .9991 & .9992 & .9993 & .9994 \\
    \cline{2-11}
.008 & PPV & .3769 & .3831 & .3892 & .3951 & .4010 & .4067 & .4123 & .4178 & .4206 \\
     & NPV & .9980 & .9981 & .9983 & .9985 & .9986 & .9988 & .9989 & .9991 & .9992 \\
    \cline{2-11}
.01 & PPV & .4310 & .4375 & .4438 & .5500 & .4560 & .4620 & .4677 & .4734 & .4762 \\
    & NPV & .9975 & .9977 & .9979 & .9981 & .9983 & .9985 & .9987 & .9989 & .9990 \\
    \cline{2-11}
.02 & PPV & .6048 & .6111 & .6172 & .6231 & .6288 & .6343 & .6397 & .6449 & .6475 \\
    & NPV & .9949 & .9953 & .9957 & .9961 & .9965 & .9969 & .9973 & .9977 & .9979 \\
    \cline{2-11}
.03 & PPV & .6988 & .7043 & .7096 & .7147 & .7197 & .7244 & .7291 & .7335 & .7357 \\
    & NPV & .9923 & .9929 & .9935 & .9941 & .9947 & .9953 & .9960 & .9966 & .9969 \\
    \cline{2-11}
.05 & PPV & .7978 & .8021 & .8061 & .8100 & .8137 & .8173 & .8208 & .8241 & .8257 \\
    & NPV & .9869 & .9879 & .9890 & .9900 & .9910 & .9921 & .9931 & .9942 & .9947 \\
    \cline{2-11}
.08 & PPV & .8671 & .8701 & .8729 & .8757 & .8783 & .8808 & .8832 & .8856 & .8867 \\
    & NPV & .9785 & .9802 & .9819 & .9836 & .9853 & .9870 & .9887 & .9904 & .9913 \\
    \cline{2-11}
.10 & PPV & .8929 & .8953 & .8978 & .9000 & .9022 & .9043 & .9063 & .9082 & .9091 \\
    & NPV & .9727 & .9748 & .9770 & .9791 & .9813 & .9834 & .9856 & .9878 & .9889 \\
    \cline{2-11}
.15 & PPV & .9298 & .9315 & .9331 & .9346 & .9361 & .9375 & .9388 & .9401 & .9408 \\
    & NPV & .9573 & .9606 & .0639 & .9672 & .9706 & .9740 & .9774 & .9808 & .9825 \\
    \cline{2-11}
.20 & PPV & .9494 & .9506 & .9518 & .9529 & .9540 & .9551 & .9560 & .9570 & .9574 \\
    & NPV & .9406 & .9451 & .9496 & .9542 & .9588 & .9635 & .9682 & .9730 & .9754 \\
\hline
\end{tabular} \\
{\footnotesize
$^{(1)}$Specificity is .99 \qquad $^{(2)}$Infection rate
}
\end{table}

\subsection{Multi-step batch testing procedure}
\label{sec:proc}

\begin{figure}
 \centerline{\includegraphics[height=2.5in]{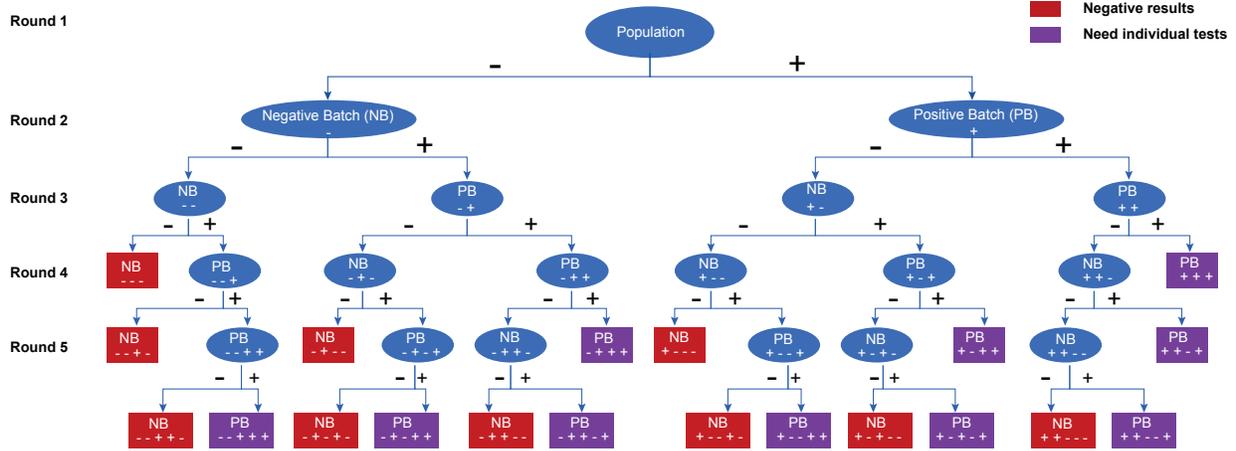}}
\caption{Proposed multi-step batch testing procedure. The oval nodes indicate intermediate nodes
  and the rectangular nodes indicate terminal nodes.
  \label{fig:tree}}
\end{figure}

More than 3 batch negatives may not be necessary in the procedure because the infection rate
substantially decreases in later rounds. Figure~\ref{fig:tree} illustrates the
proposed batch testing procedure. We will further investigate the optimal number of batch tests and
the stopping rule in this study. For batch negatives, the batch size increases substantially with
most of the subjects remaining in the next round because the infection rate decreases. For
batch positives, the subpopulation size substantially decreases in the next round. In this
group, not all the samples in the batch are infected, so we may find a sample without infection
in the next round of batch tests. This way, the number of required tests to cover the whole
population decreases significantly, and it is possible to identify all positive carriers in the
population.

We assume 15\% false negative rate and 1\% false positive rate of an individual for
a population of size 100,000 in this section. Table~\ref{tab:proc} shows the change in
the infection rate and corresponding optimal batch size throughout the process shown in
Figure~\ref{fig:tree} for a population with infection rates of .1\% and 1\%. In the first
round, the optimal batch size is 35 resulting in 2,858 batch tests when the infection rate
is .1\%, and the optimal batch size is 12 resulting in 8,334 batch tests when the infection
rate is 1\%. Note that the number of batch tests for the first round can be slightly
different from this estimation due to the trial for estimating the infection rate with
(\ref{eq:A1}) using a small subset of the population with an initial guess of the batch
size. The subsequent rounds do not require this trial because the infection rates of
subpopulations can be estimated as (\ref{eq:pi}) using the values obtained in the previous
round. For the $i$th round, the number of batch tests can be obtained by
ceiling$(N_i/n_i)$. We count only the bold-faced numbers starting the first round because
the rest are duplicates. In each of Round~2 and Round~3, the bold-faced subpopulation sizes
($N_i$) add up to 100,000 if we ignore the rounding error. In Round~4, the bold-faced
subpopulation sizes plus $N$ in the final column for already finished ($- - -$ and $+ + +$)
rows add up to 100,000. The calculation is similar for Round~5. The final two columns show
the infection rate and resulting subpopulation size in each of the terminal nodes
(rectangular frames) in Figure~\ref{fig:tree}. For each infection rate, since the
individuals in an upper half of the last column got 3 batch negatives, they do not need
individual tests. A lower half of the last column needs individual tests because they got
3 batch positives. Thus, the total number of individual tests is the sum of these 10
numbers. The total number of tests can be obtained by adding the number of batch tests in
Round~1, the number of batch tests obtained from the bold-faced numbers in the subsequent
rounds, and the number of individual tests.

\begin{table}[tbp]
 \centering
 \caption{Batch test procedure$^*$: population size 100,000, false negative rate 15\% \&
  false positive rate 1\% for individual tests, infection rates .1\% and 1\%.}
  \label{tab:proc}
\vspace{.5em}
{\footnotesize
\begin{tabular}{l|lrr|lrr|lrr|lrr|lr}
 \multicolumn{15}{c}{Infection rate$=.1\%$. The first round requires {\bf 2858} tests with batch
 size {\bf 35}.} \\
\hline
& \multicolumn{3}{c|}{Round 2} & \multicolumn{3}{c|}{Round 3} & \multicolumn{3}{c|}{Round 4} &
 \multicolumn{3}{c|}{Round 5} & \multicolumn{2}{c}{Final} \\
\hline
Tests & \multicolumn{1}{c}{$p_2^{(1)}$} & $N_2^{(2)}$ & $n_2^{(3)}$
 & \multicolumn{1}{c}{$p_3$} & $N_3$ & $n_3$ & \multicolumn{1}{c}{$p_4$} & $N_4$ & $n_4$ &
  $p_5$ & $N_5$ & $n_5$ & \multicolumn{1}{c}{$p$} & $N$ \\
\hline
$- - -$ & 2e-4 & {\bf 96109} & {\bf 88} & 2e-5 & {\bf 94047} & {\bf 224} & & & & & & & 4e-6 & 92684 \\
$- - + -$ & 2e-4 & 96109 & 88 & 2e-5 & 94047 & 224 & .001 & {\bf 1363} & {\bf 30} & & & & 2e-4 & 1302 \\
$- + - -$ & 2e-4 & 96109 & 88 & .006 & {\bf 2062} & {\bf 15} & .001 & {\bf 1888} & {\bf 35} & & & & 2e-4
   & 1814 \\
$+ - - -$ & .022 & 3891 & 8 & .004 & {\bf 3322} & {\bf 18} & 6e-4 & {\bf 3102} & {\bf 45} & & & & 1e-4
   & 3000 \\
$- - + + -$ & 2e-4 & 96109 & 88 & 2e-5 & 94047 & 224 & .001 & 1363 & 30 & .027 & {\bf 61} & {\bf 7} & .005
   & 51 \\
$- + - + -$ & 2e-4 & 96109 & 88 & .006 & 2062 & 15 & .001 & 1888 & 35 & .022 & {\bf 74} & {\bf 8} & .004
   & 63 \\
$- + + - -$ & 2e-4 & 96109 & 88 & .006 & 2062 & 15 & .062 & 175 & 5 & .012 & {\bf 133} & {\bf 10} & .002
   & 118 \\
$+ - - + -$ & .022 & 3891 & 8 & .004 & 3322 & 18 & 6e-4 & 3102 & 45 & .016 & {\bf 102} & {\bf 9} & .003 & 90 \\
$+ - + - -$ & .022 & 3891 & 8 & .004 & 3322 & 18 & .049 & 220 & 6 & .010 & {\bf 169} & {\bf 12} & .002 & 152 \\
$+ + - - -$ & .022 & 3891 & 8 & .13 & 568 & 4 & .03 & 362 & 7 & .005 & {\bf 300} & {\bf 15} & .001 & 278 \\
\hline
$+ + +$ & .022 & {\bf 3891} & {\bf 8} & .13 & {\bf 568} & {\bf 4} & & & & & & & .30 & 206 \\
$+ + - +$ & .022 & 3891 & 8 & .13 & 568 & 4 & .03 & {\bf 362} & {\bf 7} & & & & .15 & 62 \\
$+ - + +$ & .022 & 3891 & 8 & .004 & 3322 & 18 & .049 & {\bf 220} & {\bf 6} & & & & .18 & 51 \\
$- + + +$ & 2e-4 & 96109 & 88 & .006 & 2062 & 15 & .062 & {\bf 175} & {\bf 5} & & & & .22 & 42 \\
$+ + - - +$ & .022 & 3891 & 8 & .13 & 568 & 4 & .03 & 362 & 7 & .005 & 300 & 15 & .06 & 23 \\
$+ - + - +$ & .022 & 3891 & 8 & .004 & 3322 & 18 & .049 & 220 & 6 & .010 & 169 & 12 & .08 & 17 \\
$+ - - + +$ & .022 & 3891 & 8 & .004 & 3322 & 18 & 6e-4 & 3102 & 45 & .016 & 102 & 9 & .11 & 13 \\
$- + + - +$ & 2e-4 & 96109 & 88 & .006 & 2062 & 15 & .062 & 175 & 5 & .012 & 133 & 10 & .10 & 14 \\
$- + - + +$ & 2e-4 & 96109 & 88 & .006 & 2062 & 15 & .001 & 1888 & 35 & .022 & 74 & 8 & .13 & 11 \\
$- - + + +$ & 2e-4 & 96109 & 88 & 2e-5 & 94047 & 224 & .001 & 1363 & 30 & .027 & 61 & 7 & .15 & 9 \\
\hline
\end{tabular}
\vspace{.2in} \\
\begin{tabular}{l|lrr|lrr|lrr|lrr|lr}
\multicolumn{15}{c}{Infection rate$=1\%$. The first round requires {\bf 8335} tests with batch size
 {\bf 12}.} \\
\hline
& \multicolumn{3}{c|}{Round 2} & \multicolumn{3}{c|}{Round 3} & \multicolumn{3}{c|}{Round 4} &
 \multicolumn{3}{c|}{Round 5} & \multicolumn{2}{c}{Final} \\
\hline
Tests & \multicolumn{1}{c}{$p_2$} & $N_2$ & $n_2$
 & \multicolumn{1}{c}{$p_3$} & $N_3$ & $n_3$ & \multicolumn{1}{c}{$p_4$} & $N_4$ & $n_4$ &
  $p_5$ & $N_5$ & $n_5$ & \multicolumn{1}{c}{$p$} & $N$ \\
\hline
$- - -$ & .002 & {\bf 89456} & {\bf 27} & 3e-4 & {\bf 85233} & {\bf 68} & & & & & & & 4e-5 & 83107 \\
$- - + -$ & .002 & 89456 & 27 & 3e-4 & 85233 & 68 & .009 & {\bf 2126} & {\bf 12} & & & & .0015 & 1921 \\
$- + - -$ & .002 & 89456 & 27 & .03 & {\bf 4223} & {\bf 7} & .006 & {\bf 3496} & {\bf 15} & & & & .0009
   & 3229 \\
$+ - - -$ & .08 & 10544 & 5 & .017 & {\bf 7399} & {\bf 9} & .003 & {\bf 6425} & {\bf 21} & & & & .0005
   & 6033 \\
$- - + + -$ & .002 & 89456 & 27 & 3e-4 & 85233 & 68 & .009 & 2126 & 12 & .079 & {\bf 205} & {\bf 5} & .017
   & 144 \\
$- + - + -$ & .002 & 89456 & 27 & .03 & 4223 & 7 & .006 & 3496 & 15 & .061 & {\bf 267} & {\bf 5} & .012
   & 204 \\
$- + + - -$ & .002 & 89456 & 27 & .03 & 4223 & 7 & .15 & 727 & 4 & .038 & {\bf 430} & {\bf 6} & .007 & 351 \\
$+ - - + -$ & .08 & 10544 & 5 & .017 & 7399 & 9 & .003 & 6425 & 21 & .041 & {\bf 392} & {\bf 6} & .008 & 314 \\
$+ - + - -$ & .08 & 10544 & 5 & .017 & 7399 & 9 & .11 & 974 & 4 & .025 & {\bf 657} & {\bf 8} & .0044 & 550 \\
$+ + - - -$ & .08 & 10544 & 5 & .23 & 3144 & 3 & .065 & 1679 & 5 & .013 & {\bf 1262} & {\bf 10} & .0022
   & 1120 \\
\hline
$+ + +$ & .08 & {\bf 10544} & {\bf 5} & .23 & {\bf 3144} & {\bf 3} & & & & & & & .42 & 1466 \\
$+ + - +$ & .08 & 10544 & 5 & .23 & 3144 & 3 & .065 & {\bf 1679} & {\bf 5} & & & & .22 & 417 \\
$+ - + +$ & .08 & 10544 & 5 & .017 & 7399 & 9 & .11 & {\bf 974} & {\bf 4} & & & & .29 & 317 \\
$- + + +$ & .002 & 89456 & 27 & .03 & 4223 & 7 & .15 & {\bf 727} & {\bf 4} & & & & .31 & 298 \\
$+ + - - +$ & .08 & 10544 & 5 & .23 & 3144 & 3 &  .065 & 1679 & 5 & .013 & 1262 & 10 & .10 & 142 \\
$+ - + - +$ & .08 & 10544 & 5 & .017 & 7399 & 9 & .11 & 974 & 4 & .025 & 657 & 8 & .13 & 107 \\
$+ - - + +$ & .08 & 10544 & 5 & .017 & 7399 & 9 & .003 & 6425 & 21 & .041 & 392 & 6 & .18 & 78 \\
$- + + - +$ & .002 & 89456 & 27 & .03 & 4223 & 7 & .15 & 727 & 4 & .038 & 430 & 6 & .18 & 79 \\
$- + - + +$ & .002 & 89456 & 27 & .03 & 4223 & 7 & .006 & 3496 & 15 & .061 & 267 & 5 & .22 & 63 \\
$- - + + +$ & .002 & 89456 & 27 & 3e-4 & 85233 & 68 & .009 & 2126 & 12 & .079 & 205 & 5 & .23 & 60 \\
\hline
\end{tabular} \\
}
{\footnotesize
$^{(1)}$infection rate \qquad
$^{(2)}$subpopulation size \qquad
$^{(3)}$batch size
}
\end{table}
The expected number of tests to identify all positive carriers is 6,144 (5,696 batch tests and
448 individual tests) when the infection rate is .1\%, and 22,436 (19,409 batch tests and 3,027
individual tests) when the infection rate is 1\%. The sensitivity of this procedure can be
obtained from the number of individual tests with the corresponding infection rates given in
the last two columns in Table~\ref{tab:proc}. For the infection rate of .1\%, the expected number
of infected cases among the samples who received individual tests at the end of the procedure is
97.8 out of 448. With the 85\% sensitivity for individual tests, the sensitivity of this procedure
is 83.1\% since $97.8 \times .85 = 83.1$ out of 100. For the infection rate of 1\%, the
expected number of infected cases among the samples who received individual tests is 975.5 out of
3,027. The sensitivity of this procedure is 82.9\% since $975.5 \times .85 = 829$ out of 1,000.
To calculate the specificity, for the infection rate of .1\%, the expected number of uninfected
samples among the samples who received individual tests is 350.2 out of 448. With the 1\% type I
error rate for individual tests, since $350.2 \times .01 = 3.502$, the specificity of this
procedure is $100 \left[ 1-(3.502/99,900) \right]\%=99.996\%$. For the infection rate of
1\%, the expected number of uninfected samples among the samples who received individual
tests is 2,051.5 out of 3,027. Since $2,051.5 \times .01 = 20.515$, the specificity of this
procedure is $100 \left[ 1-(20.515/99,000) \right]\%=99.98\%$. These results closely match with
our simulation results given for method~(E) in Table~\ref{tab:simul} of Section~\ref{sec:simul}.

To reduce the false negative rate, we propose to conduct sequential individual tests as
follows: We conduct up to 3 tests for the same person sequentially until
a positive test occurs. Let us define $E$ as the event that an individual is infected, and
$B_i$ as the event that an individual got a positive result in the $i$th individual
test. Then the probability that a sample is tested positive in the first test is
\[ p_1 \equiv P(B_1) = P(B_1|E) P(E) + P \left( B_1 | E^C \right) P \left( E^C \right), \]
the probability that a person is tested negative and then tested positive is
\[ p_2 \equiv P \left( B_1^C \cap B_2 \right) = P \left( B_1^C \cap B_2 | E \right) P(E)
  + P \left( B_1^C \cap B_2 | E^C \right) P \left( E^C \right) \]
and a person is tested negative twice is
\[ p_3 \equiv P \left( B_1^C \cap B_2^C \right) = 1-(p_1+p_2). \]
Thus, the expected number of the individual tests for each person is
\begin{equation}
 p_1 + 2p_2+3p_3.
 \label{eq:expseq}
\end{equation}
We can obtain the expected total number of sequential individual tests by substituting the
infection rate and the subpopulation size in each of the bottom 10 rows of the last two
columns in Table~\ref{tab:proc}. For the infection rate of .1\%, the expected number of sequential
individual tests in (\ref{eq:expseq}) is 1,155, and the expected number of tests for the
whole procedure becomes 6,851. For the infection rate of 1\%, the expected number of
sequential individual tests is 7,236, and the expected number of tests for the whole
procedure is 26,645. The sensitivity of this method can be obtained by multiplying
$1-\beta^3=1-.15^3=.996625$ to the expected number of infected cases in the sequential individual
testing. Recall that the expected number of infected cases among the samples who received
individual tests was 97.8 and 975.5 for infection rates .1\% and 1\%, respectively.
For the infection rate of .1\%, the sensitivity of this method is 97.5\% since
$97.8 \times .996625 = 97.5$ out of 100. For the infection rate of 1\%, the sensitivity of
this method is 97.2\% since $975.5 \times .996625 = 972$ out of 1,000. To calculate the
specificity of this method, the probability that an uninfected sample is concluded as
positive in the sequential testing is $1-(1-\alpha)^3=1-.99^3=.029701$.
For the infection rate of .1\%, since $350.2 \times .029701 = 10.4$, the specificity of
this method is $100 \left[ 1-(10.4/99,900) \right]\%=99.99\%$. For the infection rate of
1\%, since $2,051.5 \times .029701 = 60.931$, the specificity of this method is
$100 \left[ 1-(60.931/99,000) \right]\%=99.94\%$. These results closely match with our
simulation results given for method~(F) in Table~\ref{tab:simul} of Section~\ref{sec:simul}.


\begin{table}[tbp]
 \centering
 \caption{Possible cases of the proposed batch test procedure with false negative rates.}
 \label{tab:cases}
\vspace{1em}
\begin{tabular}{c|ccccccc|ccccccc}
\hline
& \multicolumn{7}{c|}{Conclusion to the final batch:}
    & \multicolumn{7}{c}{Individual tests} \\
& \multicolumn{7}{c|}{no infection}
    & \multicolumn{7}{c}{for the final batch} \\
\hline
& & \multicolumn{5}{c}{Batch test round} & False & & \multicolumn{5}{c}{Batch test round}
  & False \\
\cline{3-7} \cline{10-14}
\#tests & Case & 1 & 2 & 3 & 4 & 5 & negative rate & Case & 1 & 2 & 3 & 4 & 5 & negative rate \\
\hline
3 & (1) & $-$ & $-$ & $-$ & & & $\beta^3$ & (11) & $+$ & $+$ & $+$ & & & $(1-\beta)^3 \beta^3$ \\
4 & (2) & $-$ & $-$ & $+$ & $-$ & & $(1-\beta) \beta^3$ &
  (12) & $+$ & $+$ & $-$ & $+$ & & $(1-\beta)^3 \beta^4$ \\
4 & (3) & $-$ & $+$ & $-$ & $-$ & & $(1-\beta) \beta^3$ &
  (13) & $+$ & $-$ & $+$ & $+$ & & $(1-\beta)^3 \beta^4$ \\
4 & (4) & $+$ & $-$ & $-$ & $-$ & & $(1-\beta) \beta^3$ &
  (14) & $-$ & $+$ & $+$ & $+$ & &  $(1-\beta)^3 \beta^4$ \\
5 & (5) & $-$ & $-$ & $+$ & $+$ & $-$ & $(1-\beta)^2 \beta^3$ &
  (15) & $+$ & $+$ & $-$ & $-$ & $+$ & $(1-\beta)^3 \beta^5$ \\
5 & (6) & $-$ & $+$ & $-$ & $+$ & $-$ & $(1-\beta)^2 \beta^3$ &
  (16) & $+$ & $-$ & $+$ & $-$ & $+$ &  $(1-\beta)^3 \beta^5$ \\
5 & (7) & $-$ & $+$ & $+$ & $-$ & $-$ & $(1-\beta)^2 \beta^3$ &
  (17) & $+$ & $-$ & $-$ & $+$ & $+$ &  $(1-\beta)^3 \beta^5$ \\
5 & (8) & $+$ & $-$ & $-$ & $+$ & $-$ & $(1-\beta)^2 \beta^3$ &
  (18) & $-$ & $+$ & $+$ & $-$ & $+$ &  $(1-\beta)^3 \beta^5$ \\
5 & (9) & $+$ & $-$ & $+$ & $-$ & $-$ & $(1-\beta)^2 \beta^3$ &
  (19) & $-$ & $+$ & $-$ & $+$ & $+$ &  $(1-\beta)^3 \beta^5$ \\
5 & (10) & $+$ & $+$ & $-$ & $-$ & $-$ & $(1-\beta)^2 \beta^3$ &
  (20) & $-$ & $-$ & $+$ & $+$ & $+$ &  $(1-\beta)^3 \beta^5$ \\
\hline
\end{tabular}
\end{table}

To further illustrate how the false negative rate is reduced by this
procedure, Table~\ref{tab:cases} lists all possible cases shown in Table~\ref{tab:proc} with the
false negative rate in each case. Among the 20 possible cases of batch
testing, the first 10 cases result in 3 batch negatives, and the second 10 cases result in 3 batch
positives. If a sample is in one of the first 10 cases, then it will be deduced as uninfected, and
if it is in one of the second 10 cases, then it requires up to 3 sequential individual tests. The
false negative rate is $\beta^3=.003375$ for Case~(1) because it has 3 batch negatives,
$(1-\beta) \beta^3=.00287$ for Cases~(2) through (4) because it has 3 batch negatives and 1 batch
positive, and $(1-\beta)^2 \beta^3=.00244$ for Cases~(5) through (10) because it has 3 batch
negatives and 2 batch positives. Since false negative occurs when a sample gets negative in all
three individual tests sequentially, $\beta^3$ is multiplied by the rates given in
Table~\ref{tab:cases} to the second 10 cases. The false negative rate is $(1-\beta)^3
\beta^3=.00207$ for Case~(11), $(1-\beta)^3 \beta^4=.00031$ for Cases~(12) through (14), and
$(1-\beta)^3 \beta^5=.000047$ for Cases~(15) through (20). An estimation of the false
negative rate of the proposed batch testing procedure can be obtained by multiplying the false negative
rate with the subpopulation size given in the last column of Table~\ref{tab:proc} in each case and
divide the sum of these 20 numbers by the population size of 100,000. The estimated false negative
rate is 3.33\% when the
infection rate is .1\%, and 3.25\% when the infection rate is 1\%. These results yield the sensitivity
of 96.67\% and 96.75\%, respectively. These estimates of the sensitivity are slightly lower and the
estimates by combining groups of 10 cases in the previous paragraph (97.5\% and 97.2\%, respectively)
are slightly higher than the simulation results for method~(F) in Table~\ref{tab:simul} given in
Section~\ref{sec:simul}, but they are within one standard deviation (1.65\% for infection rate .1\%
and .57\% for infection rate 1\%).

\section{Simulation Studies}
\label{sec:simul}

We conducted a Monte Carlo simulation study to evaluate the efficiency and
efficacy of the proposed batch testing procedure. In this simulation, we assume
that the sensitivity and specificity of individual tests are 85\% and 99\%, respectively.
A population of 100,000 people is randomly generated 100 times. The infection rates
of .1\%, 1\%, 3\%, 5\% and 10\% are chosen. Table~\ref{tab:simul} compares our
methods with conventional individual tests, single batch tests, and matrix pool tests
using various accuracy measures. We compare the accuracy measures for (A) conventional
individual tests, (B) one-step batch testing with a fixed batch size of 10,
(C) one-step batch testing with optimal batch sizes, with up to 3 sequential
individual tests for each of the positive batches, (D) parallel $12 \times 12$ matrix
pool testing (to cover the whole population, 694 $12 \times 12$ matrices and one
$8 \times 8$ matrix are used); up to 3 sequential individual tests for all the
positive intersections. Since a single matrix testing gives very low sensitivity, we
report results from three parallel matrix pool tests. (E) proposed multi-step batch
tests ending with three batch negatives or three
batch positives, with an individual test given to three batch positives,
(F) multi-step batch tests given in (E), with up to 3 sequential individual tests
for three batch positives. For (C), (E) and (F), the optimal batch sizes estimated in
Section~\ref{sec:proc} are used in the subpopulations in each step. For each of the
infection rates, the overall accuracy, sensitivity, specificity, PPV and NPV are
calculated, and the required number of tests to cover the whole population is
calculated from this simulation, and the values are averaged over the 100 repetitions.
The fixed batch size of 10 given in (B) has been used in South Korea for high-risk
facilities. For multi-step batch testing, individual tests are conducted to the
subpopulation if the infection rate exceeds 30\% in any step as mentioned in
Section~\ref{sec:intro}. Another set of simulation results for sensitivity of 75\%
and specificity of 97\% is given in the Appendix. The comparison between our methods
and the conventional individual tests, single batch testing, or matrix pool tests
gives a similar pattern to the one given in this section.

\begin{table}[tbp]
{ \centering
 \caption{Simulation results: 100 repetitions, population size 100,000, sensitivity 85\% \&
  specificity 99\% for individual tests; mean with standard deviation in parentheses.}
  \label{tab:simul}
\vspace{.5em}
{\small
\begin{tabular}{ccccccc}
\hline
$p^{(1)}$ &   & .001     & .01    & .03   & .05    & .10 \\
\hline
(A)$^a$ & Acc.$^{(2)}$  & .9899 (.0003) & .9886 (.0003) & .9859 (.0004) & .9831 (.0004) & .9760 (.0004)  \\
Indiv  & Sens.$^{(3)}$  & .8525 (.0356) & .8491 (.0105) & .8500 (.0067) & .8509 (.0051) & .8497 (.0030)  \\
Tests   & Spec.$^{(4)}$ & .9900 (.0003) & .9900 (.0003) & .9901 (.0003) & .9900 (.0003) & .9900 (.0003)  \\
        & PPV           & .0799 (.0088) & .4621 (.0107) & .7251 (.0075) & .8176 (.0054) & .9041 (.0031) \\
        & NPV           & .9998 (.0000) & .9985 (.0001) & .9954 (.0002) & .9921 (.0003) & .9834 (.0004)  \\
      & \#Tests$^{(5)}$ & 100000 (0)    & 100000 (0)    & 100000 (0)    & 100000 (0)     & 100000 (0)   \\
\hline
(B)$^b$ & Acc.  & .9995 (.0001) & .9964 (.0002) & .9897 (.0003) & .9831 (.0005) & .9675 (.0006) \\
Single  & Sens. & .7261 (.0413) & .7213 (.0163) & .7241 (.0093) & .7219 (.0075) & .7221 (.0052) \\
Batch   & Spec. & .9998 (.0000) & .9992 (.0001) & .9979 (.0002) & .9968 (.0002) & .9948 (.0003) \\
Tests   & PPV   & .8074 (.0449) & .8998 (.0101) & .9136 (.0057) & .9227 (.0039) & .9389 (.0028) \\
Fixed   & NPV   & .9997 (.0001) & .9972 (.0002) & .9915 (.0003) & .9855 (.0004) & .9699 (.0006) \\
Size 10 & \#Tests & 11841 (133) & 18981 (259) & 33108 (438) & 44693 (500) & 65671 (519) \\
\hline
(C)$^c$ & Acc. & .9987 (.0001) & .9956 (.0002) & .9911 (.0003) & .9868 (.0004) & .9784 (.0005) \\
Single & Sens. & .8420 (.0384) & .8453 (.0128) & .8464 (.0074) & .8470 (.0059) & .8466 (.0042) \\
Batch & Spec.  & .9989 (.0001) & .9971 (.0002) & .9956 (.0002) & .9941 (.0002) & .9930 (.0003) \\
Variable & PPV & .4308 (.0268) & .7460 (.0124) & .8548 (.0067) & .8835 (.0040) & .9303 (.0027) \\
Sizes & NPV    & .9998 (.0000) & .9984 (.0001) & .9952 (.0002) & .9920 (.0003) & .9832 (.0005) \\
Seq Indiv & \#Tests & 14393 (1128) & 37889 (824) & 60567 (817) & 77952 (793)  & 98341 (739) \\
Tests & B+Ind  & 2858+11535    & 8334+29555    & 14286+46281   & 16667+61285   & 25000+73341 \\
\hline
(D)$^d$ & Acc.          & .9999 (.0000) & .9994 (.0001) & .9970 (.0002) & .9939 (.0003) & .9863 (.0005) \\
Parallel  & Sens.       & .9907 (.0106) & .9897 (.0048) & .9905 (.0031) & .9901 (.0031) & .9905 (.0029) \\
$12 \times 12$  & Spec. & .9999 (.0000)  & .9995 (.0001) & .9972 (.0002) & .9941 (.0003) & .9858 (.0005) \\
Matrix   & PPV          & .9402 (.0332) & .9511 (.0077) & .9168 (.0053) & .8980 (.0041) & .8857 (.0032) \\
Tests & NPV             & 1.000 (.0000) & .9999 (.0000) & .9997 (.0001) & .9995 (.0002) & .9989 (.0003) \\
{\small Seq Indiv} & \#Tests & 50727 (342) & 56201 (550) & 80478 (971) & 112177 (1458) & 190263 (1951) \\
Test & B+Ind           & 49968+759 & 49968+6233 & 49968+30510 & 49968+62209 & 49968+140295 \\
\hline
(E)$^e$ & Acc.              & .9998 (.0000) & .9981 (.0001) & .9941 (.0003) & .9905 (.0003) & .9814 (.0005)  \\
{\small Multi-step} & Sens. & .8305 (.0337) & .8281 (.0120) & .8306 (.0066) & .8291 (.0053) & .8296 (.0041) \\
Batch & Spec.               & 1.000 (.0000) & .9998 (.0001) & .9992 (.0001) & .9990 (.0001) & .9983 (.0001) \\
Variable  & PPV             & .9503 (.0241) & .9722 (.0058) & .9693 (.0037) & .9776 (.0021) & .9815 (.0014) \\
Sizes & NPV                 & .9998 (.0000) & .9983 (.0001) & .9948 (.0002) & .9911 (.0003) & .9814 (.0005) \\
1 indiv  & \#Tests          & 6221 (159)    & 22840 (280)   & 42561 (316)   & 56089 (377) & 84012 (375)  \\
Test      & B+Ind           & 5705+516      & 19439+3401    & 31830+10731   & 41554+14535 & 58622+25390 \\
\hline
{\bf (F)$^f$ Multi-} & Acc. & .9998 (.0000) & .9990 (.0001) & .9969 (.0002) & .9958 (.0002) & .9926 (.0003) \\
{\bf step Batch} & Sens.    & .9689 (.0165) & .9700 (.0057) & .9730 (.0033) & .9723 (.0025) & .9725 (.0017) \\
{\bf Variable}   & Spec.    & .9999 (.0000) & .9993 (.0001) & .9976 (.0002) & .9970 (.0002) & .9949 (.0002) \\
{\bf Sizes} & PPV           & .8852 (.0290) & .9307 (.0085) & .9259 (.0044) & .9446 (.0028) & .9547 (.0018) \\
{\bf Seq Indiv}  & NPV      & 1.000 (.0000) & .9997 (.0001) & .9992 (.0001) & .9985 (.0001) & .9969 (.0002) \\
{\bf Tests for} & \#Tests   & 7053 (259)    & 27809 (404)   & 58428 (609)   & 75908 (600) & 116341 (718) \\
{\bf 3} $+$'s & B+Ind       & 5704+1349     & 19436+8373    & 31809+26619   & 41530+34378 & 58611+57730 \\
\hline
\end{tabular} \\
} }
{\footnotesize
$^a$Conventional individual tests \vspace{-.04in} \\
$^b$One-step batch tests with a fixed batch size of 10, individual tests for positive
 batches  \vspace{-.04in} \\
$^c$One-step batch tests with variable batch sizes with up to 3 sequential individual tests
  for positive batches \vspace{-.04in} \\
$^d$Three parallel matrix pool tests; up to 3 sequential individual tests to all positive
  intersections \vspace{-.04in} \\
$^e$Multi-step batch tests with variable optimal batch sizes; an individual test for
 3 batch positives \vspace{-.04in} \\
$^f$Multi-step batch tests; up to 3 sequential individual tests for 3 batch positives \vspace{-.04in} \\
$^{(1)}$infection rate \quad $^{(2)}$overall accuracy \quad $^{(3)}$sensitivity \quad $^{(4)}$specificity
\quad $^{(5)}$Number of required tests \vspace{-.04in} \\
$^{(6)}$number of batch tests $+$ number of individual tests
}
\end{table}

For the matrix pool testing, we also simulated conventional one-step matrix pool
test with an individual test to the positive intersection, but the sensitivity was
less than 62\% for all the infection rates. For one-step matrix pool tests with up
to 3 sequential tests to each of the positive intersection, the sensitivity was
around 72\%. For 3 parallel matrix pool tests with an individual test to all
positive intersections, the sensitivity was around 84\%. To achieve high performance
for the comparison, we report simulation results for 3 parallel matrix pool tests
with up to 3 sequential individual tests to all positive intersections in (D).
Although not reported in this paper, single-step batch testing with variable batch
sizes also was conducted. The accuracy measures were very close to those of the
fixed batch size in (B). However, the batch tests with variable batch sizes require
fewer tests.

Note that the number of required tests in the proposed method (F) is close to the
estimated number derived in Section~\ref{sec:proc}. The sensitivity is reduced in
the one-step batch tests (given in (B)) to approximately 72\% from
conventional individual tests. This is because if a batch is tested negative, then
all the samples in the batch are considered uninfected, and no further tests are
given. The results are in line with the sensitivity of 72.25\% for one-step batch
testing obtained in Section~\ref{sec:sens}. Method~(C) substantially improved the
sensitivity to almost the same level as the sensitivity of conventional individual
tests. The sensitivity of method~(E) is around 83\%, and it is improved
by method~(F) to approximately 97\%. These results show that both the multi-step
batch testing and sequential individual tests significantly improve the sensitivity.
These are in line with the numbers obtained from the model in
Section~\ref{sec:proc}. This improvement is achieved because sequential individual
tests are given to small target subpopulations obtained by our test procedure. Although
not included in Table~\ref{tab:simul}, we also conducted a simulation of multi-step
batch testing with a fixed batch size of 10 followed by up to 3 sequential individual
tests. The sensitivity of this method was almost identical to that of method~(F) because
the sensitivity does not depend on the batch size. However, method~(F) requires much
fewer tests by using the optimal batch sizes (7,053 for method~(F) vs.\ 32,118 for a
fixed batch size for infection rate .1\%, and 27,809 for method~(F) vs.\ 46,416 for a fixed
batch size for infection rate 1\%). The sensitivity of method~(D) is around 99\%. However,
the number of tests for matrix pool testing is 50,727 for the infection rate of .1\% and 56,201 for
the infection rate of 1\%. Although not included in the simulation table, we also simulated
3 parallel batch testing with a fixed batch size of 10. The sensitivity of the method was
near 85\%, and the number of tests to cover the whole population was slightly over 30,000
which is larger than that of method~(F) for infection rates of .1\% or 1\%, but fewer for
the infection rates higher than 1\%. The sensitivity of this parallel batch testing was much
lower than that of method~(F).

As mentioned in Section~\ref{sec:spec}, the specificity is significantly improved by
batch testing from conventional individual testing. The specificity of single batch
testing obtained in this simulation closely matches with the expected value given
in (\ref{eq:spec}). See also Table~\ref{tab:spec}. Although the specificity of the
batch tests decreases as the infection rate increases, the one-dimensional batch tests
(B, C, E and F) have much higher specificity than the baseline of 99\% for
all infection rates in Table~\ref{tab:simul}. However, the specificity of the matrix
pool tests decreases fast as the infection rate increases and it becomes lower than 99\%
for the infection rate of 10\%.

The improvement of the PPV by our procedure is remarkable. As discussed in
Section~\ref{sec:ppvnpv} (also shown in (A) of Table~\ref{tab:simul}), the PPV of
an individual test is 8\% and 46\% when the infection rates are .1\% and 1\%,
respectively. According to our simulation, single-step batch testing in (B)
improves this to 81\% and 90\%, respectively, and it is further
improved by multi-step batch testing given in (E) to 95\% and 97\%, respectively.
The PPV is decreased by sequential individual tests. It is 43\% and 75\%, respectively
in (C). However, it is improved to 88.5\% and 93\%, respectively in (F) by multi-step batch
testing. The PPV is 94\% and 95\%, respectively for the matrix pool tests in (D).

The overall number of batch tests from the simulation is close to the estimated
number from our model given in Section~\ref{sec:model}. For .1\% infection rate,
the number of multi-step batch tests in the simulation (methods (E) and (F)) is around
5,700 and the estimated number
is 5,696. For 1\% infection rate, the number is around 19,440 from the simulation and
19,409 from our estimation. However, the simulated number of individual tests for samples
with 3 batch positives is slightly higher than the predicted value. This is likely
due to randomness. This discrepancy does not significantly impact the overall number
of tests because the size of the subpopulation requiring individual tests is very small
(approximately 0.5\% for .1\% infection rate, and around 3\% for 1\% infection rate)
compared to the whole population. For .1\% infection rate,
for example, 516 people received individual tests in the simulation
(given in (E)), whereas the model estimates that 448 people need individual
tests. This means approximately 99.5\% of the population do not require individual
tests. The model estimate of this value is 99.552\% ($100,000-448$ out of $100,000$),
and the simulated value is 99.484\% ($100,000-516$ out of $100,000$).

Note that the overall accuracy is mostly affected by specificity because the
infection rates are low. We can incorporate geographic and demographic information
for more realistic calculation.

\section{Discussion}

The COVID-19 pandemic changed our lifestyle, seriously impacted the global economy, and
took many precious lives. To get back to normalcy, we need a rapid testing of the virus
for all the residents of each community.
Unlike other coronavirus outbreaks we experienced in the past, the disease rapidly
spreads silently by asymptomatic carriers. Since only patients with symptoms have been
getting tests, it is a challenging task to identify asymptomatic COVID-19 carriers. In
most countries including the US, some patients with symptoms could not get tests due to
the limited testing capacity. To conduct testing a broader population more efficiently,
batch testing methods have been introduced.

The South Korean Center for Disease Control \& Prevention used a single-step batch
testing for long-term care facilities with a fixed batch size of 10 for the entire
staff and patients. As seen in our simulation studies, batch testing increases the
false negative rate, although this approach can monitor high-risk groups without
symptoms by reducing the number of tests needed to cover the entire community. In this
paper, we proposed a multi-step batch testing procedure to substantially decrease the
false negative rate using a small number of test kits to completely test a large
population. The improvement of PPV from individual testing is also remarkable, and thus
our multi-step approach can be trusted for reliable results. Table~\ref{tab:proc}
shows that the proposed batch procedure is effective for a population with a low or
moderate infection rate.

Our approach will be useful for the prevention for early stages of future pandemics.
Our method is most effective for diseases with infection rates of up to 3\%. We do
not recommend this approach for highly contagious large populations with infection
rates greater than 5\%, as the prescribed number of tests becomes very large.
Shuren (2020) addressed that conventional batch testing has a higher chance
of false negative results because samples are diluted, but it works well when
there is a low prevalence of cases. Yelin et al.\ (2020) found that positive
samples can still be well observed in pools in up to 32 samples, and possibly
even 64 with additional PCR cycles. However, the detection of positives in a large
pool is possible only when viral loads are very high. In future studies, we will
investigate the presence of the dilution effect. We will also investigate
optimal stopping rules to further improve the efficiency and efficacy of the
multi-step batch testing procedure.

\section*{Acknowledgments}

Xiaolin Li and Haoran Jiang are supported in part by US Army Research Office grant
W911NF-18-10346, Xiaolin Li and Hongshik Ahn are supported by the US Army Research
Office through the equipment grant W911NF-20-10159. The authors thank Andrew Ahn for
useful discussions and giving valuable comments on a draft of this paper.

\section*{References}

\begin{singlespace}
\begin{list}{0}{\setlength{\rightmargin}{\leftmargin}}
\item[
Amemiya, C. T., Algeria-Harman, M. J., Aslanidis, C.,] Chen, C., Nikolic, J., Gingrich, J. C. and
 de Jong, P. J. (1992). A two-dimensional YAC pooling strategy for library screening via STS and
 Alu-PCR methods. {\em Nucleic Acids Research}, {\bf 25}, 2559-63.
\item[
Armend{\'a}riz, I., Ferrari, P. A., Fraiman, D. and Dawson, S. P.] (2020).
 Group testing with nested pools. https://arxiv.org/abs/2005.13650.
\item[
Barillot, E., Lacroix, B. and Cohen, D.] (1991). Theoretical analysis of library screening using
 a N-dimensional pooling strategy. {\em Nucleic Acids Research}, {\bf 19}, 6241-6247.
\item[
Behets, F., Bertozzi, S., Kasali, M.,] Kashamuka, M., Atikala, L., Brown, C., Ryder, R. W. and
 Quinn, T. C. (1990). Successful use of pooled sera to determine HIV-1 seroprevalence in Zaire
 with development of cost-efficiency models. {\em AIDS}, {\bf 4}, 737-741.
\item[
Bilder, C. R. and Tebbs, J. M.] (2009). Bias, efficiency, and agreement for group-testing
 regression models. {\em The Journal of Statistical Computation and Simulation}, {\bf 79}, 67-80.
\item[
Busch, M., Caglioti, S., Robertson, E., McAuley, J.,] Tobler, L., Kamel, H., Linnen, J.,
 Shyamala, V., Tomasulo, P. and Kleinman, S. (2005). Screening the blood supply for West Nile
 virus RNA by nucleic acid amplification testing. {\em New England Journal of Medicine},
 {\bf 353}, 460-467.
\item[
Chen, C. L. and Swallow, W. H.] (1990). Using group testing to estimate a proportion, and to test
 the binomial model. {\em Biometrics}, {\bf 46}, 1035-1046.
\item[
Chen, P., Tebbs, J. M. and Bilder, C. R.] (2009). Group testing regression models with fixed and
 random effects. {\em Biometrics}, {\bf 65}, 1270-1278.
\item[
Cheng, Y. Y.] (2020).
  Statistical methods for batch screening of input populations by stage and group in COVID-19 nucleic
  acid testing.  {\em medRxiv}. doi: 10.1101/2020.04.02.20050914.
\item[
Delaigle, A. and Meister, A.] (2011).
 Nonparametric regression analysis for group testing data.
 {\em Journal of the American Statistical Association}, {\bf 106}, 640-650.
\item[
Dorfman, R.] (1943).
 The detection of defective members of large populations. {\em The Annals of Mathematical Statistics},
 {\bf 14}, 436-440.
\item[
Fahey, J. W., Ourisson, P. J. and Degnan, F. H.] (2006). Pathogen detection, testing and
 control in fresh broccoli sprouts. {\em Nutrition Journal}, {\bf 5}, 13.
\item[
Farrington, C.] (1992). Estimating prevalence by group testing using generalized linear models.
 {\em Statistics in Medicine}, {\bf 11}, 1591-1597.
\item[
Fletcher, R. H., Fletcher, S. W. and Wagner, E. H.] (1988).
 {\em Clinical Epidemiology} (2nd ed.), Baltimore, Williams and Wilkins, pp. 53-60.
\item[
France, B., Bell, W., Chang, E. and Scholten, T.] (2015), Composite sampling approaches for
  bacillus anthracis surrogate extracted from soil. {\em PLOS One}. \\
   doi: 10.1371/journal.pone.0145799.
\item[
Gastwirth, J. L. and Hammick, P. A.] (1989). Estimation of prevalence of a rare disease,
 preserving the anonymity of the subjects by group testing: Application to estimating the
 prevalence of AIDS antibodies in blood donors. {\em Journal of Statistical Planning and
 Inference}, {\bf 22}, 15-27.
\item[
Gastwirth, J. L. and Johnson, W. O.] (1994). Screening with cost-effective quality control:
 Potential applications to HIV and drug testing. {\em Journal of the American Statistical
 Association}, {\bf 89}, 972-981.
\item[
Hardwick, J., Page, C. and Stout, Q.] (1998). Sequentially deciding between two experiments
 for estimating a common success probability. {\em Journal of the American Statistical
 Association}, {\bf 93}, 1502-1511.
\item[
Hogan, C. A., Sahoo, M. K. and Pinsky, B. A.] (2020). Sample pooling as a strategy to detect community
 transmission of SARS-CoV-2. {\em Journal of the American Medical Association}, {\bf 323(19)},
 1967-1969. doi: 10.1001/jama.2020.5445.
\item[
Hourfar, M., Jork, C., Schottstedt, V., Weber-Schehl, M., Brixner, V.,] Busch, M., \\
 Geusendam, G., Gubbe, K., Mahnhardt, C., Mayr-Wohlfar, W., Pichl, L., Roth, W.,
 Schmidt, M., Seifried, E. and Wright, D. (2008). Experience of German red cross blood donor
 services with nucleic acid testing: results of screening more than 30 million blood
 donations for human immunodeficiency virus, hepatitis C virus, and hepatitis B virus.
 {\em Transfusion}, {\bf 48}, 1558-1566.
\item[
Huang, X.] (2009). An improved test of latent-variable model misspecification in structural
 measurement error models for group testing data. {\em Statistics in Medicine}, {\bf 28},
 3316-3327.
\item[
Huang, X. and Tebbs, J. M.] (2009). On latent-variable model misspecification in structural
 measurement error models for binary response. {\em Biometrics,}, {\bf 65}, 710-718.
\item[
Hudgens, M. G. and Kim, H.-Y.] (2011). Optimal configuration of a square array group testing
 algorithm. {\em Communications in Statistics - Theory and Methods}, {\bf 40:3}, 436-448.
\item[
Hung, M. C. and Swallow, W. H.] (2000). Use of binomial group testing in tests of hypotheses
 for classification or quantitative covariables. {\em Biometrics}, {\bf 56}, 204-212.
\item[
Hwang, F. K.] (1976). Group testing with a dilution effect. {\em Biometrika}, {\bf 63}, 671-680.
\item[
Kim, H. Y., Hudgens, M. G., Dreyfuss, J. M.,] Westreich, D. J. and Pilcher, C. D. (2007).
 Comparison of group testing algorithms for case identification in the presence of test error.
 {\em Biometrics}, {\bf 63}, 1152-1163.
\item[
Kline, R. L., Brothers, T. A., Brookmeyer, R., Zeger, S.] and Quinn, T. C. (1989).
 Evaluation of human immunodeficiency virus seroprevalence in population surveys
 using pooled sera. {\em Journal of Clinical Microbiology}, {\bf 27}, 1449-1455.
\item[
Korea Center for Disease Control \& Prevention.] (2020).
 Frequently asked questions for KCDC on COVID-19 (updated on 24 April). {\em Press Release}.
 URL: https://www.cdc.go.kr/board/ \\
 board.es?mid=a30402000000\&bid=0030.
\item[
Kwak, S.] (2020).
 Korea considers testing pooled samples for vulnerable groups. {\em Korea Biomedical Review}. \\
 URL: http://www.koreabiomed.com/news/articleView.html?idxno=7966.
\item[
Lee, D. and Lee, J.] (2020).
 Testing on the move: South Korea's rapid response to the COVID-19 pandemic. {\em
 Transportation Research Interdisciplinary Perspectives}, {\bf 5}.
 doi: 10.1016/j.trip.2020.100111.
\item[
Lennon, J. T.] (2007). Diversity and metabolism of marine bacteria cultivated on dissolved DNA.
 {\em Applied and Environmental Microbiology}, {\bf 73}, 2799-2805.
\item[
Lindan, C., Mathur, M., Kumta, S., Jerajani, H., Gogate, A.,] Schachter, J. and Moncada, J.
 (2005). Utility of pooled urine specimens for detection of Chlamydia trachomatis and Neisseria
 gonorrhoeae in men attending public sexually transmitted infection clinics in Mumbai, India,
 by PCR. {\em Journal of Clinical Microbiology}, {\bf 43}, 1674-1677.
\item[
Litvak, U., Tu, X. M. and Pagano, M.] (1994). Screening for the presence of a disease by
 pooling sera samples. {\em Journal of the American Statistical Association}, {\bf 89}, 424-434.
\item[
Lohse, S., Pfuhl, T., Berk{\'o}-G{\"o}ttel, B., Rissland, J., Gei{\ss}ler, T.,] G{\"a}rtner, B.,
 Becker, S. L., Schneitler, S. and Smola, S. (2020).
 Pooling of samples for testing for SARS-CoV-2 in asymptomatic people.
 {\em The Lancet Infectious Diseases}. \\ doi: 10.1016/S1473-3099(20)30362-5.
\item[
McMahan, C. S., Tebbs, J. M. and Bilder, C. R.] (2013).
 Regression models for group testing data with pool dilution effects. {\em Biostatistics},
 {\bf 14(2)}, 284-298.
\item[
Mutesa, L., Ndishimye, P., Butera, Y., Souopgui, J., Uwineza, A.,] Rutayisire, R.,
Ndoricimpaye, E. L., Musoni, E., Rujeni, N., Nyatanyi, T., Ntagwabira, E.,
Semakula, M., Musanabaganwa, C., Nyamwasa, D., Ndashimye, M., Ujeneza, E.,
Mwikarago, I. E., Muvunyi, C. M., Mazarati, J. B., Nsanzimana, S., Turok, N. and
Ndifon, W. (2020). A pooled testing strategy for identifying SARS-CoV-2 at low
prevalence. {\em Nature}. \\
doi: org/10.1038/s41586-020-2885-5
\item[
Nagi, M. S. and Raggi, L. G.] (1972). Importance to ``airsac'' disease of water supplies
 contaminated with pathogenic escherichia coli. {\em Avian Diseases}, {\bf 16}, 718-723.
\item[
Park, D. H. and Koo, B. K.] (2020).
 S. Korea to start using pooling methods to test 10 samples at a time. {\em Hankyoreh Daily}.
 URL: http://english.hani.co.kr/arti/english\_edition/ \\
 e\_national/936568.html.
\item[
Phatarfod, R. M. and Sudbury, A.] (1994). The use of a square array scheme in blood testing.
 {\em Statistics in Medicine}, {\bf 13}, 2337-2343.
\item[
Pilcher, C., Fiscus, S., Nguyen, T., Foust, E., Wolf, L.,] Williams, D., Ashby, R.,
 O'Dowd, J., McPherson, J., Stalzer, B., Hightow, L, Miller, W., ,Eron, J., Cohen, M.
 and Leone, P. (2005). Detection of acute infections during HIV testing in North Carolina.
 {\em New England Journal of Medicine}, {\bf 352}, 1873-1883.
\item[
Shani-Narkiss, H., Gilday, O. D., Yayon, N. and Landau, I. D.] (2020).
 Efficient and practical sample pooling for high-throughput PCR diagnosis of COVID-19.
 {\em medRxiv}. \\ doi: 10.1101/2020.04.06.20052159.
\item[
Shuren, J.] (2020).
 Coronavirus (COVID-19) update: Facilitating diagnostic test availability
 for asymptomatic testing and sample pooling. {\em Press Announcement, CDRH Offices, US Food
 and Drug Administration}. URL:
 https://www.fda.gov/news-events/press-announcements/coronavirus-covid-19-update
                   -facilitating-diagnostic-test-availability \\
                   -asymptomatic-testing-and.
\item[
Stramer, S., Notari, E., Krysztof, D. and Dodd, R.] (2013). Hepatitis B.\ virus testing by \\
 minipool nucleic acid testing: does it improve blood safety? {\em Transfusion}, {\bf 53},
 2449-2458.
\item[
US Food and Drug Administration] (2020).
 Coronavirus (COVID-19) Update: FDA issues first emergency authorization for sample
 pooling in diagnostic testing.
 {\em FDA New Release}. URL:
 https://www.fda.gov/news-events/press-announcements/coronavirus-covid-19- \\
 update-fda-issues-first-emergency-authorization-sample-pooling-diagnostic.
\item[
Van, T. T., Miller, J., Warshauer, D. M., Reisdorf, E.,] Jernigan, D., Humes, R.
 and Shult, P. A. (2012). Pooling nasopharyngeal/throat swab specimens to increase
 testing capacity for influenza viruses by PCR. {\em Journal of Clinical Microbiology},
 {\bf 50}, 891-896.
\item[
Vansteelandt, S., Goetghebeur, E. and Verstraeten, T.] (2000).
Regression models for disease prevalence with diagnostic tests on pools of serum samples.
 {\em Biometrics}, {\bf 56}, 1126-1133.
\item[
Wahed, M. A., Chowdhury, D., Nermell, B., Khan, S. I., Ilias, M.,] Rahman, M.,
 Persson, L. A. and Vahter, M. (2006). A modified routine analysis of arsenic content in
 drinking-water in Bangladesh by hydride generation-atomic absorption spectrophotometry.
 {\em Journal of Health, Population and Nutrition}, {\bf 24}, 36-41.
\item[
Wang, B., Han, S., Cho, C., Han, J., Chen, Y., Lee, S.,] Galappaththy, G., Thimasam, K., \\
 Sope, M., Oo, H., Kyaw, M. and Han, E. (2014). Comparison of microscopy, nested PCR, and
 real-time PCR assays using high-throughput screening of pooling samples for diagnosis
 of malaria in asymptomatic carriers from areas of endemicity in Myanmar. {\em Journal
 of Clinical Microbiology}, {\bf 52}, 1838-1845.
\item[
Wang, D., McMahan, C., Gallagher, C. and Kulasekera, K.] (2014).
 Semiparametric group testing regression models. {\em Biometrika}, {\bf 101}, 587-598.
\item[
Warasi, M. S., Tebbs, J. M., McMahan, C. S. and Bilder, C. R.] (2016).
 Estimating the prevalence of multiple diseases from two-stage hierarchical pooling.
 {\em Statistics in Medicine}, {\bf 35(21)}, 3851-3864.
\item[
West, C. P., Montori, V. M. and Sampathkumar, P.] (2020).
 COVID-19 testing: The threat of false-negative results.
 {\em Mayo Clinic Proceedings}, {\bf 95(6)}, 1127-1129. \\
 doi: 10.1016/j.mayocp.2020.04.004.
\item[
Xiao, A. T., Tong, Y. X. and Zhang, S.] (2020).
 False-negative of RT-PCR and prolonged nucleic acid conversion in COVID-19:
 Rather than recurrence. {\em Journal of Medical Virology}. \\
 doi: 10.1002/jmv.25855.
\item[
Xie, M.] (2001). Regression analysis of group testing samples. {\em Statistics in
 Medicine}, {\bf 20}, 1957-1969.
\item[
Yang, W. and Yan, F.] (2020).
 Patients with RT-PCR-confirmed COVID-19 and normal chest CT.
 {\em Radiology}, {\bf 295(2)}, E3. doi: 10.1148/radiol.2020200702.
\item[
Yelin, I., Aharony, N., Tamar, E. S., Arogoetti, A., Messer, E., Berenbaum, E.,]
  Shafran, E., Kuzli, A., Gandali, N., Hashimshony, T., Mandel-Gutfreund, Y.,
  Halberthal, M., Geffen, Y., Szwarcwort-Cohen, M. and Kishony, R. (2020),
  Evaluation of COVID-19 RT-qPCR test in multi-sample pools.
  {\em medRxiv}. doi: 10.1101/2020.03.26.20039438.
\end{list}
\end{singlespace}

\newpage
\begin{table}
{\Large \bf Appendix \hfill}
\vspace{.1in}
{ \centering \caption{Simulation results: 100 repetitions, population size 100,000, sensitivity
75\% \& specificity 97\% for individual tests; mean with standard
deviation in parentheses.}
\label{tab:simul2}
\vspace{0.5em}
{\small
\begin{tabular}{ccccccc}
\hline
$p^{(1)}$  &  & .001 & .01  & .03  & .05  & .10 \\
\hline
(A)  & Acc.$^{(2)}$     & .9698 (.0005)  & .9678 (.0005)  & .9635 (.0006) & .9590 (.0005) & .9479 (.0007)\\
Indiv  & Sens.$^{(3)}$  & .7528 (.0427)  & .7505 (.0127) & .7498 (.0077)  & .7502 (.0062)  & .7497 (.0040) \\
Tests  & Spec.$^{(4)}$  & .9700 (.0005) & .9700 (.0005) & .9700 (.0005)  & .9700 (.0005) & .9700 (.0006)\\
 & PPV                  & .0249 (.0029)  & .2019 (.0060) & .4354 (.0065)  & .5681 (.0059)  & .7347 (.0050) \\
 & NPV                  & .9997 (.0001)  & .9974 (.0002)  & .9921 (.0003)  & .9866 (.0004)  & .9721 (.0005) \\
 & \#Tests$^{(5)}$  & 100000 (0)  & 100000 (0)  & 100000 (0)  & 100000 (0)  & 100000 (0) \\
\hline
(B)  & Acc.         & .9985 (.0002)  & .9929 (.0003)  & .9810 (.0005) & .9697 (.0006)  & .9435 (.0008)\\
Single  & Sens.     & .5582 (.0504)  & .5609 (.0146)  & .5631 (.0099)  & .5621 (.0078)  & .5633 (.0053) \\
Batch  & Spec.      & .9989 (.0001)  & .9972 (.0002)  & .9939 (.0003)  & .9911 (.0003)  & .9858 (.0004) \\
Tests  & PPV        & .3444 (.0359)  & .6717 (.0158)  & .7413 (.0079)  & .7690 (.0067)  & .8154 (.0042) \\
Fixed  & NPV        & .9995 (.0001)  & .9956 (.0002)  & .9866 (.0004)  & .9773 (.0005)  & .9531 (.0008) \\
Size 10  & \#Tests  & 13719 (186)    & 19825 (279)   & 31897 (391)    & 41845 (476)  & 59885 (494) \\
\hline
(C) & Acc.          & .9948 (.0004) & .9877 (.0005) & .9778 (.0005) & .9710 (.0007) & .9521 (.0008) \\
Single & Sens.      & .7355 (.0441) & .7377 (.0141) & .7374 (.0083) & .7381 (.0070) & .7379 (.0054) \\
Batch & Spec.       & .9951 (.0004) & .9902 (.0005) & .9853 (.0005) & .9832 (.0005) & .9759 (.0006) \\
Variable & PPV      & .1321 (.0110) & .4315 (.0092) & .6081 (.0065) & .6981 (.0061) & .7724 (.0045) \\
Sizes & NPV         & .9997 (.0001) & .9973 (.0002) & .9918 (.0003) & .9862 (.0004) & .9711 (.0007) \\
Seq Indiv & \#Tests & 19141 (1176) & 40726 (992)    & 63015 (984) & 74769 (815)  & 102424 (776) \\
Tests & B+Ind       & 2632+16509    & 7693+33033    & 12500+50515   & 16667+58102   & 20000+82424 \\
\hline
(D) & Acc.             & .9991 (.0003) & .9967 (.0003) & .9887 (.0005) & .9791 (.0007) & .9966 (.0011) \\
Parallel  & Sens.      & .9561 (.0224) & .9544 (.0092) & .9536 (.0069) & .9539 (.0076) & .9541 (.0065) \\
$12 \times 12$ & Spec. & .9991 (.0003)  & .9971 (.0003) & .9897 (.0005) & .9804 (.0007) & .9568 (.0010) \\
Matrix & PPV           & .5389 (.0878) & .7672 (.0202) & .7422 (.0083) & .7189 (.0061) & .7108 (.0048) \\
Tests & NPV            & 1.000 (.0000) & .9995 (.0001) & .9986 (.0002) & .9975 (.0004) & .9947 (.0008) \\
{\small Seq Indiv} & \#Tests & 53039 (986) & 60833 (1027) & 87098 (1414) & 118421 (1809) & 192161 (1975) \\
Test & B+Ind  & 49968+3071 & 49968+10865 & 49968+37130 & 49968+68453 & 49968+142193 \\
\hline
(E) & Acc.          & .9995 (.0001)  & .9958 (.0002)  & .9878 (.0003)  & .9803 (.0004)  & .9616 (.0006) \\
{\small Multi-step} & Sens.  & .6713 (.0485)  & .6715 (.0154)  & .6717 (.0090)  & .6784 (.0064)
  & .6785 (.0052) \\
Batch & Spec.       & .9998 (.0001)  & .9990 (.0001)  & .9975 (.0002)  & .9962 (.0002)  & .9931 (.0003) \\
Variable & PPV      & .7887 (.0444)  & .8743 (.0112)  & .8944 (.0061)  & .9039 (.0048)  & .9160 (.0031) \\
Sizes & NPV         & .9997 (.0001)  & .9967 (.0002)  & .9899 (.0003)  & .9833 (.0004)  & .9653 (.0007) \\
1 indiv & \#Tests   & 7072 (167)  & 24991 (285)  & 47494 (355)  & 60867 (374)  & 87170 (398) \\
Test & B+Ind        & 6375+697  & 20892+4099  & 36860+10634  & 44268+16599  & 57405+29765 \\
\hline
{\bf (F) Multi-}$^f$  & Acc.  & .9994 (.0001)  & .9960 (.0002)  & .9896 (.0004)  & .9840 (.0004)  & .9711 (.0006) \\
{\bf step Batch} & Sens.  & .8887 (.0289)  & .8832 (.0104)  & .8841 (.0061)  & .8904 (.0045)  & .8914 (.0034) \\
{\bf Variable}  & Spec.  & .9995 (.0001)  & .9971 (.0002)  & .9929 (.0003)  & .9889 (.0003)  & .9799 (.0005) \\
{\bf Sizes}  & PPV  & .6342 (.0346)  & .7557 (.0115)  & .7930 (.0063)  & .8083 (.0044)  & .8314 (.0036) \\
\bf{Seq Indiv}  & NPV  & .9999 (.0000)  & .9988 (.0001)  & .9964 (.0002)  & .9942 (.0002)  & .9878 (.0004) \\
\bf{Tests for} & \#Tests  & 8290 (312)  & 31393 (467)  & 63634 (627)  & 85252 (670)  & 129406 (746) \\
{\bf 3} $+$'s & B+Ind  & 6383+1906  & 20890+10502  & 36877+26757  & 44241+41011  & 57346+72061 \\
\hline
\end{tabular} \\
} }
{\footnotesize
$^{(1)}$infection rate \quad $^{(2)}$overall accuracy \quad $^{(3)}$sensitivity \quad $^{(4)}$specificity
\quad $^{(5)}$Number of required tests \vspace{-.04in} \\
$^{(6)}$number of batch tests $+$ number of individual tests
}
\end{table}

\end{document}